\documentclass[12pt]{article}

\usepackage[utf8]{inputenc}
\usepackage[english]{babel}
\usepackage{natbib}
\usepackage{amsmath,amssymb,amsfonts} 
\usepackage[dvipdfm]{graphicx}
\usepackage[dvipsnames]{xcolor}
\usepackage{latexsym}   
\usepackage{pgf,tikz}
\usepackage{geometry}
\usepackage{lineno}
\usepackage{caption}
\usepackage{graphicx}
\usepackage{booktabs}
\usepackage{array}
\usepackage[nodisplayskipstretch]{setspace}
\usepackage{hyperref}
\usepackage{multirow}
\usepackage[mathscr]{euscript}
\usepackage{pifont}
\usepackage{tcolorbox}
\usepackage{pdflscape}

\hypersetup{
    colorlinks = true,
	citecolor=Blue,
	linkcolor=RedViolet
}

\newcommand{\tick}{\ding{51}}
\newcommand{\cross}{\ding{55}}

\tcbset{colback = black!3!white,
        colframe = black!75!white,
        coltitle=white,
        colbacktitle  = black!75!white,
        fonttitle= \bfseries}       

\begin{document}

\title{\textbf{Cross validation for model selection: \\a primer with examples from ecology}}
\author{Luke A. Yates$^1$, Zach Aandahl$^1$, Shane A. Richards$^1$, and Barry W. Brook$^1$\\[7mm]
{\small $^1$School of Natural Sciences, University of Tasmania, Hobart, Tasmania, Australia}}
\date{{$\,$}\\ \begin{flushleft}
\end{flushleft}}
\maketitle
\begin{spacing}{1.0}  
\captionsetup[figure]{font={stretch=1.0}} 
\vspace{-34mm}
\begin{center}
\textbf{Abstract}\\[2mm]
\end{center} 	
The growing use of model-selection principles in ecology for statistical inference is underpinned by information criteria (IC) and cross-validation (CV) techniques. Although IC techniques, such as Akaike's Information Criterion, have been historically more popular in ecology, CV is a versatile and increasingly used alternative. CV uses data splitting to estimate model scores based on (out-of-sample) predictive performance, which can be used even when it is not possible to derive a likelihood (e.g., machine learning) or count parameters precisely (e.g., mixed-effects models and penalised regression). Here we provide a primer to understanding and applying CV in ecology. We review commonly applied variants of CV, including approximate methods, and make recommendations for their use based on the statistical context. We explain some important---but often overlooked---technical aspects of CV, such as bias correction, estimation uncertainty, score selection, and parsimonious selection rules. We also address misconceptions (and truths) about impediments to the use of CV, including computational cost and ease of implementation, and clarify the relationship between CV and information-theoretic approaches to model selection. The paper includes two ecological case studies---both from modelling contexts wherein traditional IC are difficult to apply---which illustrate the application of the techniques. The first is a classification task based on either parametric or machine-learning models; the second is a non-linear hierarchical regression problem focused on selecting allometric growth models. We conclude that CV-based model selection should be widely applied to ecological analyses, because of its robust estimation properties and the broad range of situations for which it is applicable. In particular, we recommend using leave-one-out (LOO) or approximate LOO CV to minimise bias, or otherwise $K$-fold CV using bias correction if $K<10$. To mitigate overfitting, we recommend calibrated selection via the modified one-standard-error rule which accounts for the predominant cause of overfitting: score-estimation uncertainty.


\newpage

\section*{Introduction}

The use of model-selection principles to compare and select amongst candidate working hypotheses has slowly, but surely, overtaken null hypothesis statistical testing (NHST) as the de facto standard in ecology and evolution in recent decades \citep{aho2014,Tredennick2021}. A substantial literature has emerged which identifies known shortcomings of NHST such as an a priori bias towards the null model and the potential for spurious claims of significance. These shortcomings have underscored the theoretical advantages of alternatives approaches to NHST: in particular, information-theoretic (IT) model selection \citep{Johnson1999}. For the implementation of the IT approach in ecology, the influential work of \citet{Burnham2002} advocated for the use of information criteria, while the adoption of cross validation (CV) has progressed more slowly, despite its compatibility with IT principles and its broader range of applicability \citep{Hooten2015}. Based on data splitting, CV is a practical and theoretically rigorous framework for model selection, particularly where estimation of the relative predictive merits of a set of models forms the basis of model comparison and the subsequent selection amongst candidate model structures and variables. 

To compare and ultimately select models from a set of candidates, a measure of model performance must be chosen. A model score is a measure of the expected performance of each model under prediction to new data, where performance is quantified by a loss function such as squared error or log density. The choice of loss function determines the associated discrepancy, score and theoretical properties of the selection framework; for example, log-density loss corresponds to Kullback-Leibler discrepancy, for which the (relative) expected value is the score estimated in IT model selection \citep{Burnham2002}. For models without a density (e.g., machine learning), squared error or other loss functions such as classification error can be used \citep{Hastie2009}. Having selected a loss function that is appropriate to the predictive task at hand, the expected performance or \emph{score} must be estimated for each candidate model.

Models scores cannot be determined exactly because they are random variables defined in terms of the true, but usually unknown, distribution of the data \citep{Richards2005}. There are two main ways to estimate model scores: (1) to compute the within-sample loss and add a bias-correction to adjust for the optimism of the within-sample estimate, or (2) to simulate prediction to new data using data splitting/resampling. Information criteria such as the Akaike Information Criteria \citep[AIC,][]{Akaike1973} and its many variants are examples of the former, whereas CV and certain bootstrapping techniques belong to the latter. Whichever method is applied, there is also need to quantify the uncertainty in the estimated score as a means to mitigate overfitting; for example, using the modified one-standard-error rule \citep{Yates2021}.

CV estimates model scores by systematically splitting the original data set into one or (usually) many pairs of training and test sets. The candidate models are fit to the training data and their performance (i.e., the loss) is evaluated on the test data. Despite its simplicity, CV is a robust and low-bias score estimator which can be used with a wide range of loss functions including, but not limited to, (IT) log-density loss \citep{Zhang2008}. In contrast to most information criteria, whose validity imposes strong constraints on the data and the models, CV is non-parametric, it avoids the need for precise or explicit parameter counting, it is applicable to non-likelihood-based models, and it can accommodate hierarchical and/or autocorrelative structures by using appropriately structured data splitting \citep{Burman1994}. 

Despite its many merits, implementing CV for model selection usually requires more time and effort than using information criteria (assuming the latter is available and valid), and this has contributed to its relatively low uptake in ecology for model selection. However, modern computing capacities, approximate CV techniques, and the increasing availability of user-friendly software can often mitigate this historical concern. Another potential impediment to the use of CV in ecology is that choosing amongst the available CV variants, such as $K$-fold CV and leave-one-out (LOO) CV, and their associated options, such as the number of data splits, bias correction and stratification, can be daunting, especially as many of the results guiding implementation are buried in the dedicated statistical literature. 

Here we provide an accessible and comprehensive primer on using and understanding CV for model selection, with a focus on ecological problems. We review commonly used CV techniques and their associated options, drawing on results from the statistics literature to give clear recommendations on CV strategies (with particular emphasis for ecologists), including adjustments to manage computational demands while accounting for bias and mitigating overfitting. In particular, we recommend using LOO as a general purpose, low-bias CV technique whenever it is available and computationally achievable, and otherwise approximate LOO or bias-corrected $K$-fold, where CV estimates and their uncertainty are used in conjunction with calibrated selection rules to mitigate potential overfitting. We give an overview of some useful model scores to be used in combination with CV, with advice for score selection based on theoretical properties and the modelling contexts. The paper includes two worked examples. The first is a variable selection problem in the context of binary classification using both logistic regression and random forest (decision tree) methods. The second concerns selection amongst a set of non-linear, hierarchical allometric growth models with differing fixed and random effects structures.

\begin{small}
\setlength\extrarowheight{1pt}
\renewcommand{\arraystretch}{1}
\begin{center}
\begin{table}[!t]
  \begin{center}
  \caption{\label{tab:glossary} Glossary of terms}
  \resizebox{\textwidth}{!}{%
    \begin{tabular}[!b]{ll}
		\toprule
		\toprule
		Term & Definition\\
		\midrule
		BPI & Bayesian projective inference: a variable-selection method for valid inference\\
		calibration & quantification of the significance of expected score differences (e.g., to determine\\
		& $\,\,$when the performance of two models is comparable) \\
		confusion matrix & a matrix summarising the predictive performance of a binary classifier\\
		CV & cross validation: the use of data splitting to estimate predictive performance\\
		discrepancy & 
		the difference between the expected losses of the true and an approximating model\\  
		hyperparameter & a parameter that governs model fitting or a parameter of a prior distribution\\
		IC & information criterion: a within-sample estimator of Kullback-Leibler discrepancy (KLD)\\
		IT & information-theoretic: based on the principles of information theory\\
		KLD & Kullback-Leibler discrepancy: discrepancy based on log-density (information) loss\\
		LASSO & least absolute shrinkage and selection operator: a type of penalised regression that can \\ & $\,$ lead to rejection of weak predictors\\
		loss function & 
		a function that numerically quantifies the predictive performance of a model\\
		LOO/LOGO & leave-one-out/leave-one-group-out: types of data-splitting schemes\\
		MCC & Matthew's correlation coefficient: a confusion-matrix metric\\
		metric & (of a confusion matrix) a statistic based on the entries of a confusion matrix\\
		objective function & 
		a function that is optimised when fitting a model (e.g., log-likelihood)\\
		OSE rule (modifed) & one-standard error rule: a selection rule calibrated by score-estimation uncertainty \\
		overfitting & the inclusion of spurious predictors in a selected model, leading to imprecision\\
		penalisation & regularisation via addition of a complexity penalty to the objective function\\
		regularisation & 
		a statistical method to control/constrain the effective complexity of a model\\
		score & 
		out-of-sample loss, averaged over test and training data\\
		tuning & using the data to determine the value of a hyperparameter (e.g., using CV)\\
		TSS & true skill statistic: a confusion-matrix metric\\
		underfitting & the failure to include important predictors in a selected model, leading to bias\\
		\bottomrule
	    \end{tabular}
	    }
		\end{center}
\end{table}
\end{center}

\end{small}


\section*{The model-selection process: a brief overview}

Model selection uses the available data to compare and select amongst a set of candidate models for the typical purpose of inference and/or prediction. The set of models correspond to multiple-working hypotheses or different candidate processes that generate the data. These models differ in their specification, such as complexity (e.g., the number of variables included and/or regularisation methods), the functional relationship between variables (e.g., variable interactions or alternative non-linear functional dependencies), structure (e.g., alternative grouping or hierarchical structures), and the choice of the probability distributions. There are many technical terms commonly used with model selection; Table \ref{tab:glossary} contains a glossary of technical terms used in this paper.

\begin{figure}[t]
	\centering
		\includegraphics[width=1\textwidth]{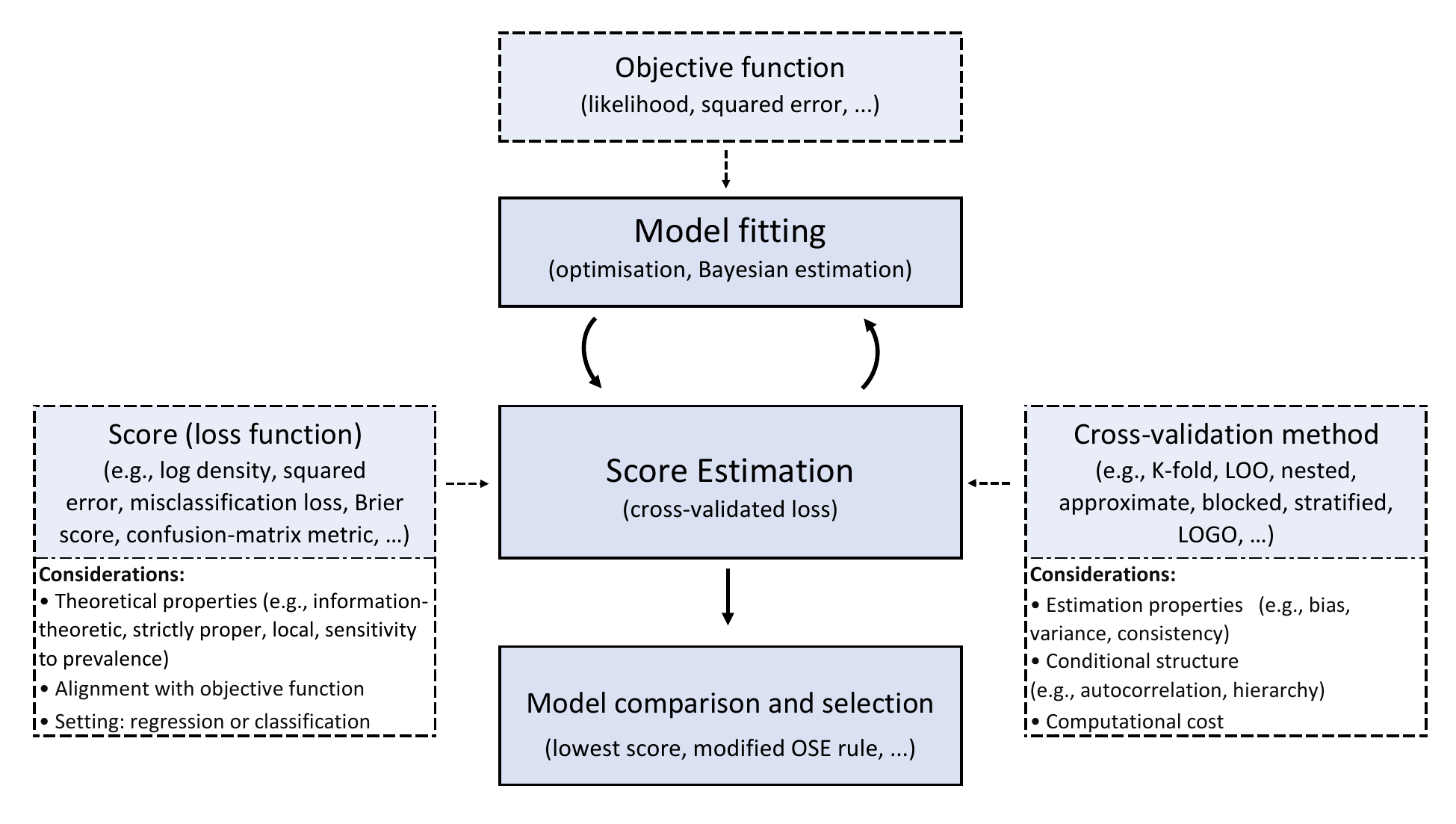}
	\caption{\label{fig:overview} Overview of the model fitting and model selection process. Given data and a set of candidate models, the process proceeds from fitting and score estimation, to comparison and selection. Models are fit with respect to a chosen objective function, and score estimation is based on the selection of both a loss function and an estimation method---application of the latter usually requires multiple fits of each model. Statistical summaries of the estimated scores (e.g., means, standard deviations, and correlations) are used in conjunction with a chosen selection rule to compare and possibly select a preferred model.}
\end{figure}

For a given set of candidate models, selection is based on estimation of the predictive performance (score) of each model, where performance is quantified by a chosen loss function and estimated using CV or in some cases, information criteria. Although the selection is based on the estimated model scores, selection rules to mitigate overfitting, or a desire for interpretable models, can lead to the preferential selection of a simpler model whose is performance is slightly worse (but usually comparable) to the best-performing model. The CV model-selection process is summarised in Figure \ref{fig:overview}. Although the predominant focus in this paper is the use of CV to select a single model (we touch briefly on model expansion), a candidate model should only be selected for inference or prediction if it passes model checking (e.g., generates plausible data under simulation and across the desired range of inference or prediction); see \emph{Model Selection} for further discussion.

\section*{Model scores}
\textbf{Loss functions, discrepancies, and scores}\\[2mm]
Fundamental to model selection is the \emph{a priori} selection of a suitable loss function $L$ to quantify predictive performance. In the regression setting, two commonly used functions are:
\begin{align*}
&L(y,\hat y)=(y - \hat y)^2  \qquad\qquad &\mbox{(squared error)}\\[2mm]
&L(y,\hat y)= \mbox{log }p(y\mid \hat y) &\mbox{(log density)}
\end{align*}
where $y$ and $\hat y$ denote observed and fitted responses, respectively, and $p$ is the model likelihood, if available. Associated with a given loss function is a discrepancy and a score, defined as follows:
\begin{align}
&\mbox{\textbf{Loss}}\quad & \displaystyle{L_m} &= \,\,L(y,\hat y_m)\label{eq:loss}\\[3mm]
&\mbox{\textbf{Discrepancy}} & D_m &= \mathrm{E}_y[L(y, y_{m^*})] - \mathrm{E}_y[L(y,\hat y_m)]\label{eq:discrepancy}\\[3mm]
&\mbox{\textbf{Score}} &S_m &= \mathrm{E}_x \mathrm{E}_y[L(y,\hat y_{m(x)})]\label{eq:score}
\end{align}
where $m = 1,..,M$ indexes candidate models, $m^*$ is the true (but usually unknown) data-generating model, and all expectations are taken over the distribution of the (multi-dimensional) data. The score is the expected discrepancy (up to an additive constant) or the double-expected loss, averaging over the randomness in both the training data $x$ and the representative test data $y$. When $L$ is log density, the corresponding discrepancy is the Kullback-Leibler Discrepancy (KLD), and the score (relative expected KLD) is the usual quantity estimated in information-theoretic model selection (e.g., AIC or CV).

The first term in the discrepancy is an unknown, but fixed constant common to all models, and can therefore be ignored when assessing the relative discrepancy of candidate models. Despite this simplification, it is generally not possible to estimate $D_m$, since the distribution of the data remains unknown. In practice, given a finite data set, the double-expected loss (i.e., the score) is much more amenable to statistical analysis \citep{Hastie2009}. For this reason it is the score, not the discrepancy, that forms the basis of model selection.

\textbf{Choosing a score}\\[2mm]
What options do we have for loss functions and their associated scores? Moreover, how do we make an appropriate choice? First, the choice of score must reflect the modelling problem at hand, taking account of the cost and benefits of differing predictive performance as well as the objective functions used to fit candidate models \citep{Vehtari2017, Gneiting2011}. For likelihood-based estimation, including both optimisation and posterior densities, log density is a common choice as it is information-theoretic and it coincides with the objective function of model fitting (i.e., [maximum] log-likelihood). Squared error coincides with the objective of least-squares regression and it is equivalent to log density for homoscedastic Gaussian models. For classification problems, common choices are log loss, the Brier score, the spherical score, and misclassification (zero-one) loss \citep{Gneiting2007}, in addition to metrics derived from the entries of the confusion matrix, such as $\kappa$, $F_1$, MCC, TSS, and many more \citep{Chicco2021, Allouche2006} (see Table \ref{tab:scores} for a summary of some common scores and Box 1 for an overview of confusion matrices and associated metrics).

In addition to alignment with the objective function, other considerations when choosing a score are the theoretical properties it possesses; the most important of these being \emph{propriety} and \emph{locality} \citep{Bernardo1979}. A score is proper if it is optimised by the true data-generating distribution, and it is strictly proper if it is uniquely optimised by this distribution; thus proper scores reward predictions which are closer to the ``truth". A score is local if it depends only on the predicted density at the observed response (i.e., $p(y\!\mid\!\hat{y})$) and not, for example, on observations that could have been obtained, but were not, or on the difference between observed and predicted outcomes (e.g., $f(y - \hat{y})$ for a binary response with $\hat{y} = 0$ or $1$).  

The class of scores associated with the Bregman divergences \citep{Zhang2008}, which includes log density, squared error, and misclassification loss are all proper scores; however, the latter is neither strictly proper nor local \citep{Gneiting2007}. The non-locality, or distance insensitivity, of misclassification loss is easily demonstrated for a binary response; for example, the predicted probabilities $p = 0.6$ and $p=0.9$ attain the same score (i.e. $0$, no loss) for the response $y=1$, using the classification threshold $c = 0.5$, despite the latter being much closer in probability. Notably, the mean absolute error $|y - \hat y|$ is not a proper score (see \citet{Gneiting2007} for an example).
%
%
\newcolumntype{S}{>{\raggedright\arraybackslash}m{13mm}}
\newcolumntype{M}{>{\raggedright\arraybackslash}m{3cm}}
\setlength\extrarowheight{10pt}
\renewcommand{\arraystretch}{1}
\begin{center}
\begin{table}[!t]
  \begin{center}
  \caption{\label{tab:scores} Summary of common scores and associated loss functions}
  \resizebox{\textwidth}{!}{%
    \begin{tabular}[!b]{>{\raggedright}p{30mm}>{\raggedright}p{37mm}l>{\centering}cc}
    & Name of Score & Loss function & (Strictly)${}^f$ Proper & Local\\
		\toprule
      \multirow{3}{*}{ \rotatebox[origin=c]{0}{\textbf{Regression}}} & {\small log density}
      & $\mathrm{log}\,p(y\mid\hat y)$
      & (\tick) & \tick \\
      & { \small {mean squared error}}& $(y-\hat{y})^2$ & (\tick) & $\mbox{\cross}^a$\\ 
      & { \small {mean absolute error}}& $|y-\hat{y}|$ & \cross & \cross\\ 
      \midrule
      \multirow{5.5}{*}{\rotatebox[origin=c]{0}{\textbf{Classification${}^b$}}} & { {log loss}}& $\mathrm{log}\,p_j$ & (\tick) & \tick \\ 
      & {\small {Brier (quadratic)} }& $\sum_{k\in\Omega}(I(j=k)-p_k)^2$& (\tick) & $\mbox{\cross}^c$\\
            & {\small {Spherical} }& $p_j/\sum_{k\in\Omega}(p_k)^2$& (\tick) & $\mbox{\cross}^c$\\
      & {\small {Misclassification loss${}^d$} }& $I(y\neq \hat y_c)$& \tick & \cross\\
      & {\small {Confusion-matrix metric${}^{d\,e}$} }& $f(M_{c})$& \cross & \cross\\
			\bottomrule 
\multicolumn{5}{l}{\hspace{-5mm} ${}^a$}{\footnotesize local for homoskedastic Guassian errors.     \hfill}\\[-3mm]
\multicolumn{5}{l}{\hspace{-5mm} ${}^b$}{\footnotesize $p_k = p(k\mid\hat y)$ denotes  the predictive probability of class $k\in\Omega$, where $\Omega$ indexes all possible classes.   \hfill}\\[-3mm]
\multicolumn{5}{l}{\hspace{-5mm} $\,\,\,$}{\footnotesize $p_j = p(y\mid \hat y)$ is the predictive probability of the observed class $y = j$.\hfill}\\[-3mm]
\multicolumn{5}{l}{\hspace{-5mm} $\,\,\,$}{\footnotesize $I(x)$ is the indicator function, returning 1 if $x$ is true and 0 otherwise. \hfill}\\[-3mm]
\multicolumn{5}{l}{\hspace{-5mm} \footnotesize ${}^c$ local for $|\Omega|=2$, i.e., binary classification\hfill}\\[-3mm]
\multicolumn{5}{l}{\hspace{-5mm} \footnotesize ${}^d$ the subscript $c$ denotes the (tuneable) threshold for the binary case $\Omega = \{0,1\}$, such that $\hat y_c = I(p_1>c)$\hfill}\\[-3mm]
\multicolumn{5}{l}{\hspace{-5mm} \footnotesize ${}^e$ a summary of common metrics based on the confusion matrix $M=M_c$ is provided in Box 1\hfill}\\[-3mm]
	\multicolumn{5}{l}{\hspace{-5mm} \footnotesize ${}^f$ bracketed check marks denote strict propriety.}
	    \end{tabular}
	    }
		\end{center}
\end{table}
\end{center}
\vspace{-8mm}

For likelihood-based regression models, log density is the recommended loss function because it is strictly proper, information-theoretic, and accommodates a broad class of modelling structures, including robust regression techniques (see Box 2). For classification problems, log loss is recommended when the properties of strict propriety and locality are deemed important, otherwise MCC or TSS are examples of general-purpose metrics to be used when all of the entries in the confusion matrix entries are needed to characterise the costs and rewards of predictive performance (see Box 1). These two metrics are identified because they are not sensitive to class prevalence; however, there are a plethora of options available, including simple misclassification loss, and we recommend further investigation for specific applications \citep{Luque2019}.


\begin{tcolorbox}[title=Box 1. Confusion matrices and metrics, sidebyside, float, floatplacement=p!, bottom = 0mm]
%
%
\textbf{Confusion matrix}\\[2mm] A confusion matrix summarises the predictive performance of a binary (two-class) classifier. Labelling one class positive and the other negative, the matrix entries are the counts of the four prediction outcomes: true positives (TP), false positives (FP), false negatives (FN), and true negatives (TN). 
\vspace{-5mm}
\setlength\extrarowheight{2pt}
\renewcommand{\arraystretch}{1}
\begin{center}
\begin{tabular}{ l  c  c  c}
& \multicolumn{2}{c}{\textit{\footnotesize True Class}} &\\
 \cline{2-3}
 & {\footnotesize \textbf{Positive}} & {\footnotesize \textbf{Negative}} & \\ 
    {\footnotesize \textbf{Positive}} & TP & FP & \multirow{2}{*}{{\rotatebox[origin=c]{0}{\footnotesize \shortstack{\emph{Predicted} \\[1mm] \emph{Class}}}}}\\  
 {\footnotesize \textbf{Negative}} & FN & TN  &  \\
 \cline{2-3}
\end{tabular}
\end{center}
\vspace{5mm}
\textbf{Tunable threshold}\\[2mm]
For models that predict the (positive) class probability $p$, rather than a dichotomous outcome, a threshold $c$ is applied such that the predicted class is positive if ${p>c}$, else negative. The threshold can be treated as a hyperparameter of the model and tuned to maximise a selected metric using CV---nested CV should be used when model selection is preceeded by hyperparameter tuning (see \emph{Nested CV}).\\[2mm]
\textbf{Metrics}\\[2mm]
To use the confusion matrix for model comparison or parameter tuning it must be summarised as a scalar statistic or metric. Although confusion-matrix metrics are not usually strictly proper scores, they are a flexible class of scores which can be tailored to application-specific needs, such as accounting for
%
\tcblower
class imbalance in the data and asymmetric cost weighting of the prediction outcomes (e.g., when a FN is more costly than a FP). The literature contains a plethora of existing metrics to choose from and we present here the definitions of some that are commonly used:
\vspace{-5mm}
\begin{footnotesize}
\begin{align*}
\begin{array}{lcl}
\mathrm{{Accuracy}}&=&\frac{TP+TN}{TP + TN + FP + FN}\qquad\qquad\qquad\qquad\\[-2mm]
\mathrm{Sensitivity} &=& \frac{TP}{TP+FN}\\[-2mm]
\mathrm{Specificity} &=& \frac{TN}{TN+FP}
\end{array}
\end{align*}
\vspace{-9mm}
\begin{align*}
\begin{array}{lcl}
F_1 &=& \frac{2TP}{2TP + FP + FN}\\[-2mm]
\kappa &=& \frac {2 \cdot (TP \cdot TN {-} FP \cdot FN)}{(TP+FP)(FP+TN)+(TP+FN)(FN+TN)}\\[-2mm]
\mathrm{TSS} &=& \mathrm{sensitivity + specificity} - 1\\[-2mm]
\mathrm{MCC} &=& \frac{TP\cdot TN - FP\cdot FN}{\sqrt{(TP+FP)(TP+FN)(TN+FP)(TN+FN)}}
\end{array}
\end{align*}
\end{footnotesize}
\noindent The true skill statistic \citep[TSS,][]{Allouche2006} and Matthew's correlation coefficient \citep[MCC,][]{Matthews1975} are particularly useful because they are not sensitive to class imbalance. Recent studies suggest that MCC is more truthful and informative than $\kappa$, $F_1$, accuracy, and even the strictly proper Brier score \citep{Chicco2021}.
\vspace{2mm}

\textbf{Estimation using CV}\\[2mm]
Confusion-matrix metrics can be estimated using CV by populating the matrix with the aggregate test outcomes of a single $K$-fold iteration (see \emph{Cross-validation techniques}). Repeated $K$-fold CV can be used to estimate the sampling variability of the metric where $K$ must be less than $n$ since LOO has only one unique split. We use repeated 10-fold CV in the scat classification example to estimate MCC. 
%
\end{tcolorbox}


\section*{Cross-validation techniques}
\textbf{What is cross validation?}\\[2mm]
Cross validation is the application of data splitting to estimate the predictive performance of one or more candidate models. CV works by fitting each model to a subset of the available data (the \emph{training} set) and then comparing the models' predictive capacities (loss) on the remaining portion of the data (the \emph{test} set). To improve the estimate, the splitting procedure is usually iterated by systematically selecting different subsets of data and summarising the overall predictive performance across iterations \citep{Arlot2010}. Despite its conceptual simplicity, CV is a theoretically rigorous method to estimate the score \eqref{eq:score} associated with a given loss function \eqref{eq:loss} \citep{Zhang2008}. For instance, it is well known that leave-one-out (LOO) CV is asymptomatically equivalent to AIC \citep{Stone1977}; indeed both of these are estimates of expected KLD. 

\textbf{Data-splitting schemes}\\[2mm]
There are many variants of CV, each characterised by different data-splitting schemes. A commonly used scheme is $K$-fold, where the available data is split into $K$ (approximately) equal-sized subsets (i.e.,``folds'') which generates $K$ distinct pairs of training/test sets, obtained by removing one fold at a time (the test set) from the full data set. This scheme can be applied once, for a single initial split (ordinary $K$-fold), or it can be repeated many times for different splits (repeated $K$-fold). The score estimate for a single $K$-fold CV is
\begin{equation}\label{eq:kfold}
\widehat{S}_{K}=\frac{1}{n}\sum_{i=1}^n L(y_i, \hat{y}_i^{-k[i]}),
\end{equation}
where $i=1,...,n$ indexes the data points, $k=1,...,K$ indexes the folds, and the superscript {\small$ -k[i]$} indicates that the training set for the fitted value excluded the fold containing the $i$th data point. 

An alternative to $K$-fold is leave-$d$-out, which involves the repeated removal of $d$ (randomly selected) test points. For a sufficiently large number of iterations, the mean leave-$d$-out estimate approaches the repeated $K$-fold estimate for $d\approx n/K$; however, $K$-fold is preferable as it guarantees of balanced draw of samples and has lower variance (see Box 3) \citep{Burman1989}. An important limiting case of both $K$-fold and leave-$d$-out is LOO, which is equivalent to $K$-fold for $K=n$. 

\textbf{Ordinary, stratified, and blocked CV}\\[2mm]
For a given splitting structure, the assignment of data points to test and training sets can be uniformly random, which is typical, or it can depend on the values of one or more categorical variables (\emph{stratified} CV), or the assignment can be determined by the spatial or temporal `distance' between training and test points (\emph{blocked} CV). 

Stratified CV is generally applied to latent-variable (or random-effects) models to balance group membership (i.e., the proportion of data within each group level) across training sets or to leave-one-group-out (LOGO) of the training set entirely. These two alternatives correspond to conditional-likelihood (i.e., prediction to existing group levels where latent effects are treated as [regularised] model parameters) and marginal-likelihood (i.e., prediction to new group levels where latent effects are `integrated out') foci, respectively \citep{Fang2011, Merkle2019}. We explore these two foci further in the Pinfish example (see \emph{Examples}), where we apply both LOO and LOGO to a set of non-linear hierarchical models. 

Blocked CV omits training points within a certain distance of the test data to account for spatial or temporal autocorrelation. It is important to use blocked CV when spatial and/or temporal correlation is visible in the model residuals (e.g., in a spatial correlogram) \citep{Roberts2017, Fletcher2018}. These structured data-assignment schemes ensure that the test data are conditionally---with respect to the model---independent (or at least conditionally exchangeable), a pre-requisite for cross-validated score estimates to be valid \citep[][Chapter 5]{BDA3}. 

\textbf{Bias-corrected CV}\\[2mm]
Cross-validated score estimates are generally biased upwards (i.e., the expected loss is overestimated) due to the training set being necessarily smaller than the full data set, but the bias is easily corrected \citep{Arlot2010}. The bias reduces as the size of training set increases, such that $K=n$ (i.e., LOO) has the minimum bias of all $K$ values in $K$-fold CV. Indeed, the bias of LOO is usually negligible and for $K\geq10$, the bias is often small enough that correction is not needed \citep{Hastie2009}. It is advisable, when possible, to check this assertion by computing and comparing the bias correction for the least- and most- complex models. 
A bias-correction term can be estimated, without any additional model fits, using the method of \citet{Burman1989}. For $K$-fold CV, the pointwise bias correction $\kappa_i$ is the difference between the within-sample loss $L(y_i, \hat{y}_i)$ and the average loss of the predictions from the $K$ training folds:
\begin{align}\label{eq:kappai}
\kappa_i = L(y_i, \hat{y}_i) - \frac{1}{K}\sum_{k=1}^K L(y_i, \hat{y}_i^{-k}).
\end{align}
The bias-corrected score estimate is $\widehat{S}_{K^*} = \widehat{S}_{K} + \kappa$, where $\kappa = \frac1n\sum \kappa_i$. Using  \eqref{eq:kfold} and \eqref{eq:kappai}, $\widehat{S}_{K^*}$ can be expressed as a pointwise sum which facilitates bias-corrected estimation of the score variance for use with calibrated model selection (see \emph{Model selection} for details). For log-density loss, $\widehat{S}_{K^*}$ can be used to estimate the bias-corrected effective number of parameters (see Box 2). For scores which are not evaluated pointwise (e.g., those based on the confusion matrix), it is generally not possible to estimate a bias-correction term using \eqref{eq:kappai}. 

Bias correction is important in model selection because score-estimation bias increases with model complexity and more complex models can be over-penalised. Failure to correct for this complexity-dependant bias results in a complexity penalty which undermines the interpretation of the selected loss function and the associated discrepancy and score; for example, model comparisons based on predictive log density are a poor approximation to IT comparisons if the expected loss estimates have a complexity-dependent bias. Although the inclusion of complexity-dependent bias is sometimes used as a strategy to mitigate overfitting \citep{Cawley2010}, we recommend using minimal biased estimates, to retain score interpretation, instead applying calibrated selection principles to account for the predominant cause of overfitting: score-estimation uncertainty (see \emph{Model Selection} for further discussion).

\textbf{Nested CV}\\[2mm]
In addition to discrete model selection, CV is commonly used to tune continuous model hyperparameters such as regularisation parameters (see \emph{Regularisation, parsimony and all that}) \citep{Hastie2009}. In this application, the cross-validated score estimate is computed across a range of fixed values of the hyperparameter to determine the value which optimises the estimate---calibrated selection rules can also be applied (see \emph{Mitigating overfitting using calibrated selection rules}). When performing model selection using CV, it is important that all model hyperparameters are tuned to each training set using an additional inner layer of CV; this is called nested CV \citep{Cawley2010}. Although it can be computationally expensive, nested CV is necessary to mitigate overfitting and place all models on an equal footing. We provide an illustration of nested CV in the scat classification example. 

 \begin{tcolorbox}[title=Box 2. Important definitions and concepts in cross validation,float, floatplacement=t!]

\vspace{1mm}
\textbf{Effective number of parameters:} As useful feature of CV is that it permits a direct estimate of the effective number of parameters $p_{\mathrm{{CV}}} $ when the loss is log density:%
\begin{equation}\label{eq:p_loo}
p_{\mathrm{{CV}}} = \ell_{\mathrm{{WS}}} - \ell_{\mathrm{{CV}}},
\end{equation}
where $\ell_{\mathrm{{WS}}} = \sum_{i=1}^n{\mathrm{log}\,p (y_i\mid y)}$ and $\ell_{\mathrm{{CV}}} = \sum_{i=1}^n{\mathrm{log}\,p (y_i\mid y^{-k[i]})}$ are the within-sample and cross-validated log likelihood of the data, respectively \citep{BDA3}. This can be used to define and thereby compare the complexities of hierarchical or regularised regression models where parameter shrinkage reduces the effective number of parameters to below the count of the model parameters.   \\[2mm]
\textbf{Robust regression}: Squared-error and Gaussian-based loss objective functions are well known to be sensitive to outliers due to the squares of large residuals generating points of high leverage \citep{Davison1997}. For parameter estimation, this sensitivity can be addressed using robust regression methods such as the use of generalised probability distributions in lieu of their classical counterparts \citep{BDA3}. For example, Student-$t$ and negative binomial distributions are robust alternatives to Gaussian and Poisson distributions, respectively. For robust model scores, log density loss applied to robust regression models is an alternative to the often used mean absolute error (MAE). The latter, while insensitive to outliers, is not a strictly proper score, and corresponds to median rather than mean performance.\\[2mm]
\textbf{Consistency}: A model-selection mechanism is consistent if it selects the true data-generating model with probability 1 as $n\rightarrow \infty$, given that the true model is in the candidate set (called an $M$-closed setting). Consistent selection can be achieved with leave-$d$-out CV by allocating the majority of the data, subject to certain asymptotic properties, to the test set \citep{Shao1993}; a suggested value is: 
\begin{align*}
 d_{\mathrm{c}} = n(1 - (log(n) - 1)^{-1}).   
\end{align*}
Viewed as an estimate of expected KLD, fitting the model to less than half of the data (e.g., $d_c = 73$ for $n=100$) introduces a large complexity-dependent bias comparable to the bias of the consistent Bayesian Information Criterion (BIC) estimator. In ecology, the true model is almost certainly not in the candidate set (called an $M$-open setting), thus the use of consistent selectors such as BIC or leave-$d_{\mathrm{c}}$-out is generally inappropriate. 

\end{tcolorbox}


\textbf{Approximate CV}\\[2mm]
Approximate CV techniques provide an alternative to data splitting (with repeats) and multiple model fits, instead typically requiring just a single fit using all of the available data.
Most approximate CV methods express the LOO score estimate as a weighted sum of the pointwise within-sample loss values, where the weights are a function of the estimated leverage $h_i$ of each data point. For linear regression with squared error loss, the leverage can be computed analytically from the design matrix $X$ (for which the element $X_{ij}$ is the value of the $j^{\mathrm{th}}$ predictor [or the indicator value of a corresponding factor level] for the $i^\mathrm{th}$ response), permitting exact computation of the LOO estimate for MSE \citep{Davison1997}:
\begin{align}\label{eq:hatloo}
\widehat{S}_{\mathrm{LOO}}=\frac{1}{n}\sum_{i=1}^n\left(\frac{y_i-\hat{y}_i}{1-h_i}\right)^2
\end{align}
where $h_i = H_{ii}$ are the diagonal elements of the hat matrix $H = X(X^TX)^{-1}X^T$ (calculated in R using the function \texttt{hatvalues}). A generalisation of \eqref{eq:hatloo} proposed by \citet{Zhang2008} computes approximate LOO estimates for the loss functions associated the entire class of Bregman divergences  (e.g., log density; see \emph{Choosing a score}); however, we are unaware of any software implementation of these generalised formulas.

In a Bayesian setting, the LOO estimate of the predictive log density can be approximated using Pareto-smoothed importance sampling (PSIS); the R package \texttt{loo} provides an easy-to-use implementation \citep{Vehtari2017}. The PSIS-LOO method generates a set of pointwise LOO estimates together with a pointwise diagnostic to determine the validity of each estimate---invalid estimates, if any, must be computed exactly (i.e., requiring additional model fits). The class of models to which the PSIS-LOO method can be applied has recently been extended to include multivariate and time-series models where data splitting is not always feasible due to the modelled covariance structure of the data \citep{Burkner2021}. We demonstrate the use of PSIS-LOO in the pinfish example.

\textbf{Choosing a CV technique}\\[2mm]
When choosing a CV method, there is often a trade-off between the statistical properties (e.g., the bias and variance) and the computational cost of the estimator. In terms of statistical properties, LOO is generally the gold standard (see Box 3), and in many instances it is practical to compute LOO exactly using $n$ model fits. Often, this computation takes just a few minutes using a modern multicore processor with parallel implementation as the CV algorithm belongs to the class of so-called `embarrassingly' parallel problems \citep{Deng2012}. For model scores which do not permit pointwise score evaluation (e.g., scores based on confusion-matrix metrics), repeated $K$-fold CV may be preferable to LOO so that the sampling variability can be estimated across repetitions. An overview of common cross-validation techniques is provided in Table \ref{tab:cv_summary}.
%
%
\newcolumntype{L}{>{\raggedright\arraybackslash}m{11.7cm}}
\newcolumntype{M}{>{\raggedright\arraybackslash}m{2.4cm}}
\setlength\extrarowheight{5pt}
\renewcommand{\arraystretch}{1}
\begin{center}
\begin{table}[p!]
  \begin{center}
  	    \caption{\label{tab:cv_summary}Summary of statistical results and recommendations for common cross-validation techniques. Further details and references are provided in the main text.}
  \resizebox{\textwidth}{!}{%
    \begin{tabular}[!b]{m{2.6cm} M L} 
	\toprule[1.2pt]
	Data-splitting scheme & Estimation method & Recommendations and statistical results\\
	\midrule[1.2pt]
    \multirow{9}{*}{\large$K$-fold} & { \small {biased} }& { \footnotesize 
    For log-density regression, analytic results suggest $K\geq10$ is required to ensure that the bias and variance are close to optimal. Similar advice, based on simulation studies, is given for classification problems (see Box 3).}\\
    & { \small {bias-corrected}} & {\footnotesize Bias-corrected estimates are recommended for $K<10$ and blocked CV, when many test data are omitted. Calculation of the corrections does not increase computational cost.}\\ 
    & { \small {repeated}} & {\footnotesize Useful for estimating sampling variability when score estimates do not factor into a pointwise sum (e.g., confusion-matrix metrics). $K$-fold CV repeated $R$ times has higher bias and variance than a single iteration of $(R\!\times\!K)$-fold CV.}\\
    & { \small {stratified}} & {\footnotesize Used for grouped data to balance group membership across folds in $K$-fold CV. Not applicable or meaningful to LOO.}\\
    \midrule
    \multirow{3.5}{*}{\large LOO} & {\small {exact}} & {\footnotesize The preferred method when computationally achievable (and is asymptotically equivalent to AIC). Bias is neglible. For linear regression, a computational shortcut permits an analytic solution with a single model fit. } \\ 
    & {\small {approximate} }& {\footnotesize An excellent alternative to exact LOO, especially for slow-fitting models. Various methods exist, some of which provide diagnostics to assess validity of the approximated estimate.} \\
    \midrule
    \multirow{3.5}{*}{\large leave-$d$-out} & {\small $d\simeq \displaystyle{\frac{n}{K}}$} & {\footnotesize Equivalent to $K$-fold for a large number of repetitions, but otherwise, unlike $K$-fold, it does not guarantee a balanced draw of test samples. Computationally inefficient.} \\
    & {\small $d>\displaystyle{\frac{n}{2}}$} & {\footnotesize The only method permitting the test sample size to exceed half of the data. This can be used for consistent selection (akin to BIC), which is generally not appropriate in ecology (see Box 2).} \\
    \midrule
    Blocked & &{\footnotesize Recommended when model residuals are spatially or temporally autocorrelated. Block size depends on the strength of correlation as a function of distance or time. Bias correction is sometimes required for large blocks.}\\
    \midrule
    LOGO & &{\footnotesize Used for grouped data to assess model performance by predictive performance to new group levels. More than one group can be left out to reduce the number of model fits.}\\
	\bottomrule[1.5pt]
	    \end{tabular}
	    }
		\end{center}
\end{table}
\end{center}
%
%
For slow fitting models (e.g., some machine-learning implementations like neural nets, or hierarchical Bayesian models), $K$-fold or approximate CV are often necessary. In terms of minimising bias and variance, it is better use of computational resources to perform a single estimate, setting $K$ as large as possible, rather averaging repeated estimates using smaller $K$ values \citep{Burman1989}, but see Box 3 for a more nuanced discussion. For the application of blocked CV, bias correction is often required, even for LOO, since the case deletions reduce the size of the training set \citep{Burman1994}; detailed recommendations and examples of blocked CV in ecology can found in the review by \citet{Roberts2017}. For latent-effects models, the choice between stratified or LOGO-based data splitting depends on the predictive goals of the analysis (i.e., prediction to existing or to new group levels). We provide a detailed comparison of these two prediction foci in the pinfish example; for further discussion, examples, and comparison with Bayesian information criteria, see \citet{Merkle2019}.

In the frequentist setting, approximate CV methods only exist for model classes that are generally fast to fit anyway, such as generalised linear models. These methods are useful for large data sets and/or nested CV applications with tuned hyperparameters, however $K$-fold is the often the only available option for slow fitting models such as generalised least squares models with correlation structures. In the Bayesian setting, the approximate method PSIS-LOO can be applied to wide variety of model classes, including complex multivariate models with correlated residual structures for which the likelihood is generally difficult to factor into a pointwise product \citep{Burkner2021}. The availability of rapidly computed CV estimates makes Bayesian methods very attractive for use with the hierarchical and multivariate classes of models which are increasingly common in ecology \citep{Bolker2009}.%


\section*{Model selection}
The selection of the best-performing model from a set of candidates is often framed as a bias-variance trade-off with respect to model complexity (see Box 3 for elaboration), where the optimal trade-off is achieved asymptotically by selecting the model with the best estimated score \citep{Hastie2009}. However, the final selection of a preferred model must take into account both the modelling goals (e.g., inference and/or prediction) and model-selection uncertainty \citep{Tredennick2021}. When the primary goal is inference, a small amount of predictive performance may be surrendered in favour of increased model interpretability. When selecting the best-scoring model (e.g., when the primary goal is prediction), the probability of overfitting increases with score-estimation uncertainty, leading to sub-optimal selection and the possible inclusion of spurious effects. In this section, we review techniques for using CV-estimated model scores to select a preferred model, taking account of the modelling goals and score-estimation uncertainty.  

\textbf{Mitigating overfitting using calibrated selection rules}\\[2mm]
A common concern regarding the use of information criteria and CV in IT model selection is their proclivity to overfit when selecting amongst candidate models \citep{Vehtari2017}. This is a valid concern, however, the predominant cause of overfitting is not the choice of score or criterion, provided the score is an unbiased estimate of relative expected Kullback-Leibler discrepancy, but instead a failure to account for the uncertainty associated with score estimation \citep{Yates2021}. Indeed, selecting the best-scoring model can lead to overfitting due to random variation alone. Overfitting due to estimation uncertainty applies to all types of loss functions, not only IT log-density loss.

An effective strategy to mitigate overfitting due to estimation uncertainty is to select the least-complex candidate model whose performance is deemed `comparable' to best scoring model \citep{Piironen2020}. To quantify the notion of comparable, the set of cross-validated loss values (e.g., the pointwise losses using LOO-CV) can be used to estimate the sampling variability of the score estimates which, in turn, can be used to visualise and suggest nominal performance thresholds for the possible selection of a simpler model. The (original) one-standard-error (OSE) rule, proposed by \citet{Breiman1984}, is an example of this approach, where the performance threshold is simply the standard error of the best score, $\sigma_{\mathrm{best}}$ (i.e., a simpler model is selected when its score estimate is within $\sigma_{\mathrm{best}}$ of the best score). However, this threshold can be problematic as it fails to account for the covariance of the scores (e.g., all models will predict poorly to a given outlier); thus overestimating the relative variation between models, which can lead to underfitting.

Our recent modification of the OSE rule addresses this issue, using the pairwise correlation coefficient of the best score with each alternative score (denoted $\rho_{\mathrm{best},m}$) to define the following correlation-adjusted standard errors, indexed by model $m$ \citep{Yates2021}:
\begin{eqnarray}\label{eq:se_mod}
\sigma_m^{{\mathrm{adj}}} \equiv\sigma_{\mathrm{best}}\sqrt{1-\,\rho_{\mathrm{best},m}}\,\,\mbox{.}
\end{eqnarray}
To apply the modified OSE rule, $\sigma^{\mathrm{adj}}_m$ is added as an error bar to a plot of the score estimates of each model, where $\sigma^{\mathrm{adj}}_{\mathrm{best}}=0$ since $\rho_{\mathrm{best},\mathrm{best}}=1$. The variance terms can be estimated using a normal approximation of the distribution of the loss values due to either CV or the (non-parametric) bootstrap. The model selected by the rule is the least complex model whose adjusted error interval includes the estimate of best scoring model (e.g., see Figures \ref{fig:scat_modsel_cv}, \ref{fig:pinfish_panel}A, and \ref{fig:pinfish_panel}B). In some instances, the model with the best score coincides with the model selected by the modified OSE rule, providing assurance that the best scoring model is not overfit. We illustrate the use of the modified OSE rule in both of the examples.

Another measure of uncertainty that appears in the recent model-selection literature is the standard error of the difference between the best score and each alternative model \citep{Piironen2020}:
\begin{eqnarray}\label{eq:se_diff}
\sigma^{\mathrm{diff}}_m &=& \sqrt{\mathrm{Var}(\Delta S_m)}\\
&=& \sigma_m^2 + \sigma_{\mathrm{best}}^2 - 2\rho_{\mathrm{best},m}\sigma_m\sigma_{\mathrm{best}}\nonumber
\end{eqnarray}
where $\Delta S_m = S_{\mathrm{best}} - S_m$. Estimates of $\Delta S_m$ sample the null-hypothesis distribution that the $m^\mathrm{th}$ model does not improve over the best scoring model; thus, $\sigma^{\mathrm{diff}}_m$ can be used to calculate probabilities related to pairwise model comparisons. The correlation-adjusted $\sigma_m^{{\mathrm{adj}}}$ is closely related to $\sigma^{\mathrm{diff}}_m$. Indeed, the former is obtained from a generalisation of the latter, subject to certain additional constraints. When used as a performance threshold in a modified OSE rule, the two thresholds will often select the same model since both account for correlation in the score estimates (see \citet{Yates2021} for further discussion). Finally, when the data set is small ($n<100$) or performance estimates of the best models are similar (i.e., those with $\Delta \widehat{S} < 4$), \citet{sivula2020} have shown that both the bootstrap and the normal approximation provide unreliable estimates of the standard error for log-density loss. However, in the context of mitigating overfitting, these estimation issues are most likely benign, since the simpler model is favoured as a consequence. 
 
Estimated performance thresholds such as $\sigma_m^{{\mathrm{adj}}}$ are a useful guide for selecting a preferred model, but we caution against their indiscriminate use without appropriate model checking and the use of expert knowledge to assess the calibration of the selection rule. Indeed, estimates of the correlation of model scores can also be used to asses the merits of model enlargement \citep{BDA3, Garthwaite2010}. For example, when a subset of the best-scoring models have similar (mean) score estimates but low correlation, then each model is capturing independent aspects of the true data-generating process which suggests that a new enlarged model, comprising structural elements from each of these models, may have superior performance. 

\textbf{Regularisation, parsimony and all that}\\[2mm]
Regularisation includes a large array of statistical methods for controlling/constraining the complexity of a model. Indeed, the use of calibrated selection rules, and variable-selection approaches more generally, can be viewed as a type of discrete regularisation, where a subset of parameters in the global model are set to zero, thereby reducing model complexity with the goal of improved prediction and/or improved inferential properties. This discrete approach can be contrasted with continuous regularisation techniques that use penalised regression such as LASSO \citep{Zou2005}, non-uniform or sparsifying priors for Bayesian approaches \citep{Piironen2017hs}, or the inclusion of hierarchical structures. All of these techniques constrain parameter estimates, shrinking them towards zero (or a distributional mean), but without necessarily removing them from the (global) model. For a detailed summary of regularisation methods and their interpretation, see \citet{Hooten2015}.

For continuous approaches, the effective reduction in complexity is governed by one or more regularisation parameters which can be estimated from the data---Bayesian priors are a natural exception, unless adopting the empirical Bayes approach. For hierarchical approaches, the associated regularisation parameters are implicitly estimated as they are a function of the estimated scale parameters. In penalised regression, CV is typically used to tune the regularisation parameters for optimal predictive performance with respect to a chosen score (see also Box 2 \emph{Nested CV}). For example, in elastic net regression, the regularised parameter estimates $\widehat{\boldsymbol\theta} =(\widehat\theta_1,\widehat\theta_2,...,\widehat\theta_p)$ are those minimising the penalised function \citep{Zou2005}:
\begin{equation}
\label{eq:elastic_net}
f(\mathbf{x};\mathbf{\boldsymbol{\widehat\theta}}) + \lambda\left(\alpha||\boldsymbol{\widehat\theta}||^1 + \frac{(1-\alpha)}{2}||\boldsymbol{\widehat\theta}||^2\right),
\end{equation}
where $f$ is the objective function (usually mean squared error or negative log density), and $||\boldsymbol\theta||^1 = \sum_{j=1}^p |\theta_j|$ and $||\boldsymbol\theta||^2 = \sum_{j=1}^p \theta_j^2$ are the $L_1$- and $L_2$-norm of the vector of model parameters, respectively. The regularisation parameter $\lambda$ determines the strength of the penalty, implementing a trade-off between the size of the model's parameter estimates (the shrinkage or effective complexity) and the minimised value of the unconstrained objective function $f$. The hyperparameter $\alpha$  ($0\leq\alpha\leq1$) indexes a family of regularised models for which the extreme values $\alpha = 1$ and $\alpha=0$ correspond to LASSO (least absolute shrinkage and selection operator) and ridge regression, respectively. Both $\alpha$ and $\lambda$ can be tuned using CV. A further useful property of elastic net regression (in addition to the regularisation of model complexity) is its remarkable tolerance to the inclusion of correlated variables via the parameter-grouping effect of the penalisation strategy which alleviates the deleterious effects of multicollinearity on estimation \citep{Hastie2015, Dormann2013}. We illustrate the use of LASSO, ridge, and elastic net regularisation in the scat classification example.  

Continuous regularisation approaches such as ridge regression generally lead to superior predictive performance relative to discrete variable selection \citep{Zou2005}. However, the latter is often preferred when the inclusion of fewer variables incurs a small (but acceptable) loss in predictive power but either makes the model more interpretable or eases the logistical cost of future data collection \citep{Afrabandpey2020}. 

Some regularisation techniques (e.g., the lasso and sparsifying priors) can be viewed as both continuous---in the sense of providing constrained parameter estimation---and discrete---in that a subset of the model parameters can be shrunk all the way to zero, or at least very close to zero, leading ultimately to their removal. A recently developed method that makes use of both continuous regularisation and discrete selection is Bayesian projective inference \citep[BPI,][]{Piironen2020}. The first step in BPI is to specify a reference model which contains all available variables and a sufficiently complex structure (e.g., hierarchical levels and/or smoothing splines) to act as an effective proxy or emulator for the data-generating mechanism. The reference model can be almost any type of model, provided steps such as regularisation are taken to mitigate overfitting. The reference model should be the best (or at least close to the best) predictive model available, providing an approximate upper limit for predictive performance and a benchmark against which to compare the performance of less complex, but more interpretable alternatives. The second step in BPI is to fit the candidate set of (interpretable) models to the predicted response of the reference model (i.e., fitting-to-the-fit, rather than fitting to the observed response). This step is usually combined with CV, where the reference model is fit to each training set, the submodels are fit using maximum-likelihood estimation to the training-set reference predictions, and model performance is estimated in the usual way by computing the mean out-of-sample loss with respect to the observed test data. In practice, approximate PSIS-LOO can usually be used, obviating the need for multiple fits. When the number of candidate variables is large, thereby prohibiting exhaustive comparison of all possible discrete alternatives, forward step selection can be implemented for which the projective approach mitigates against selection stochasticity which can be problematic in classical implementations of variable step selection. The third step is discrete model selection, where selection rules such as the modified OSE rule can be applied using the reference-model score as the best score. Finally, having made a selection, the preferred model is fit to repeated draws of the reference posterior, thereby projecting the uncertainty of the posterior reference distribution on to the subspace of the simpler model. Although BPI is a complex method, the R package \text{projpred} provides an easy-to-use implementation for a wide range of model types including generalised linear mixed models and generalised additive models \citep{projpred2020}. We illustrate the application of BPI in the scat classification example.  
%
%
\begin{tcolorbox}[title= Box 3. Bias-variance trade-off, floatplacement=!p,float, fontupper = \small]

\textbf{Choosing the best predictive model}: In model selection, the problem of finding the `best' predictive model is often viewed as a bias-variance trade-off: to find the `sweet spot' between underfitting (i.e., low complexity models with high bias and low variance) and overfitting (i.e., high complexity models with low bias and high variance) \citep{Hastie2009}. For squared-error loss, this trade-off is made explicit by the decomposition:\\[-4mm]
\begin{align*}
\mathrm{E}[(\hat y_m - y_{m^*})^2] &= \left(\mathrm{E}[y_{m^*}]-\mathrm{E}[\hat y_m]\right)^2 + \mbox{Var}[y_{m^*}-\hat y_m]
+ \mbox{Var}[\varepsilon]\\
&= \mbox{bias}^2 + \mbox{variance} +  \mbox{irreducible error},
\end{align*}
where typical trajectories of the bias${}^2$, variance and expected loss, as a function of model complexity, are shown in the following figure.
\begin{center}\includegraphics[width=0.4\textwidth]{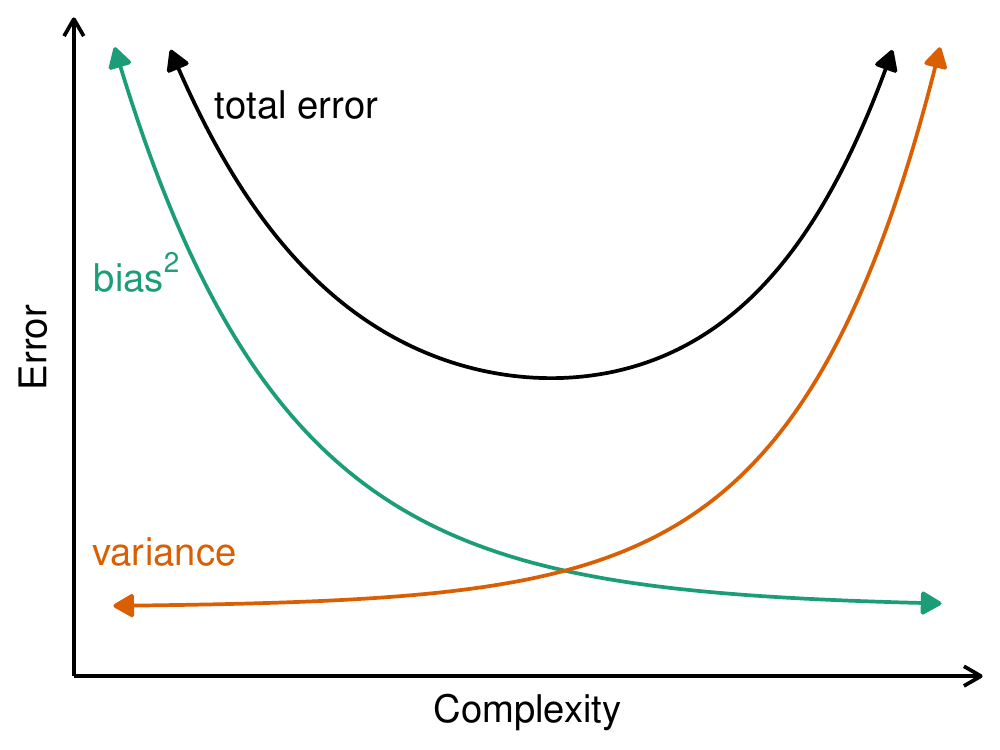}\\
\end{center}
\textbf{Choosing $K$ in $K$-fold CV}: 
It is sometimes claimed that there is a bias-variance trade-off when selecting the value of $K$ in $K$-fold CV \citep{James2013}, however the statistical literature tells a more nuanced story. Although the bias of $K$-fold CV as a score estimator is always reduced by increasing $K$, it is difficult to make universal statements about a bias-variance trade-off because the effect of $K$ on the variance depends on the estimation setting (e.g., objective function or score choice) as well as the stability of the training algorithm (e.g., model sensitivity)\citep{Arlot2010}. For example, $K=n$ is known to have the lowest bias and the lowest variance of all $K$ values in linear regression \citep{Burman1989}--this analytic (and asymptotic) result is conjectured to be true in other regression settings \citep{Arlot2010}. In contrast, on the basis of simulation studies, it is often recommended to use a much smaller $K=5$ or $K=10$, especially for classification problems \citep{Cawley2010,Hastie2009}. \\[3mm]
Even when conclusive statements can be made about the relationship between the variance and the choice of $K$, the variance of the score is not necessarily the correct quantity to examine \citep{Breiman1992}. \citet{Arlot2016} suggest that the variance of the score differences $\mathrm{Var}(S_{K,m} - S_{K,m'})$ is a more important quantity (to minimise) since model selection ultimately concerns model comparisons not score estimation per se. For least-squares density estimation, \citet{Arlot2016} show that $\mathrm{Var}(S_{K,m} - S_{K,m'})$ reduces with $K$, however close-to-optimal values are attained for $K\geq10$, affirming the existing advice (for this setting at least) to choose $K=10$ as a \emph{minimum} value when computational cost prohibits the use of LOO.
\end{tcolorbox}



\section*{Examples}

In this section we use two biological case studies to illustrate many of the CV-based model-selection approaches described in this paper. To accompany these analyses, we provide an online code repository for reproducing the complete workflow, including: data preparation, model fitting, model selection, and plot creation (see \emph{code availability}). The code provides a natural platform upon which other users can readily develop their own custom code to use CV in their statistical learning problems.

\subsection*{Scat classification}

\textbf{Dataset}\\[2mm]
Name (R package): \texttt{scat} (\texttt{caret})\\
Description: Data on animal feces in coastal California. The data consist of DNA verified species designations as well as fields related to the time and place of the collection and the morphology of the scat itself.\\[3mm]
%
\textbf{Analysis}\\[2mm]
The aim of the analysis is to predict the biological family (\emph{felid} or \emph{canid}) for each scat observation, based on eight morphological characteristics, the scat location, and the carbon-to-nitrogen ratio; see supplementary materials and the original data publication for further details concerning the data \citep{Reid2015}.

\subsubsection*{Part 1: logistic regression with MCC score}

\begin{figure}[t!]
	\centering
		\includegraphics[width=0.5\textwidth]{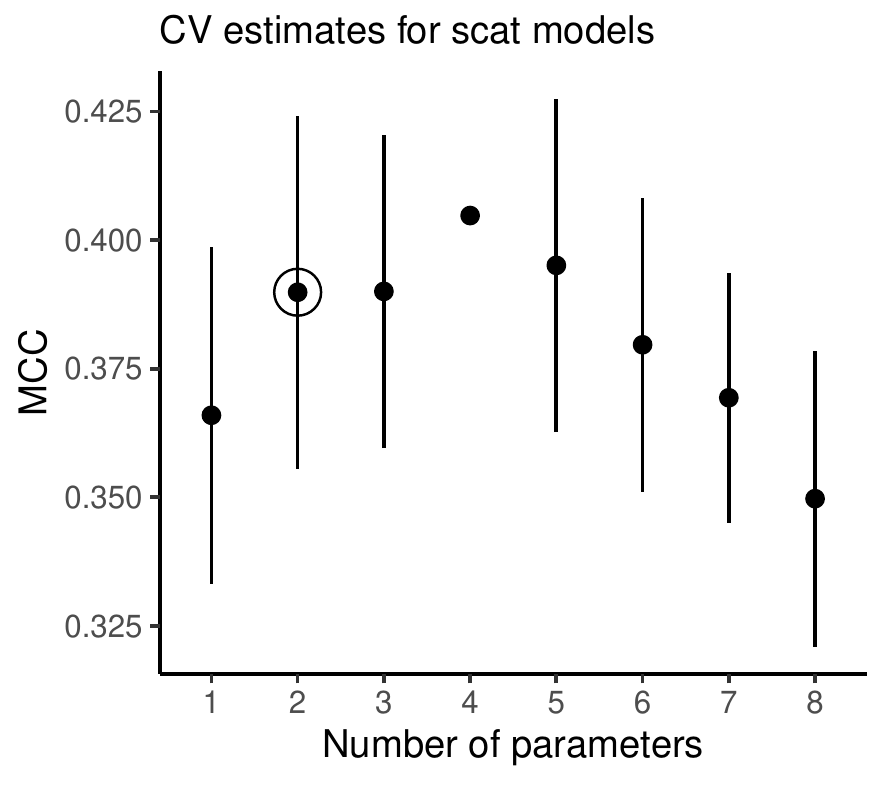}
	\caption{\label{fig:scat_modsel_cv} Comparison of logistic models using repeated 10-fold CV to estimate MCC. The dots and bars depict the mean MCC estimate and the modified standard error \eqref{eq:se_mod}, respectively. After applying the modified OSE rule, the selected model comprises two predictors: {\sf carbon-nitrogen ratio} and the {\sf number of scat pieces}.}
\end{figure}

We begin in a parametric setting, using logistic regression to model the binary class probabilities. We fit generalised linear models using maximum likelihood, but compare model performance to select variables using estimates of Matthew's correlation coefficient (MCC), as described in Box 1. Setting the probability threshold to 0.5 (threshold tuning via nested CV is possible but omitted here to simplify the exposition), we use 10-fold CV, repeated 50 times, to generate a MCC estimate for each model for each repetition. Figure \ref{fig:scat_modsel_cv} shows the mean MCC estimate and modified standard error \eqref{eq:se_mod} of the highest scoring model for each level of model complexity using all combinations of the 10 predictors (1024 models in total). To mitigate overfitting, we applied the modified OSE rule which suggests selection of the two-parameter model comprising the predictors {\sf carbon-nitrogen ratio} and the {\sf number of scat pieces}; the one-predictor model comprising only {\sf carbon-nitrogen ratio} is close in score and should also be considered. The mean model scores and estimation uncertainty for the top 10\% of all the models are shown in Figure \ref{fig:scat_modsel_top10}. The model fitting took 220 seconds using parallel processing with 40 cores.

\begin{figure}[t!]
	\centering
		\includegraphics[width=1\textwidth]{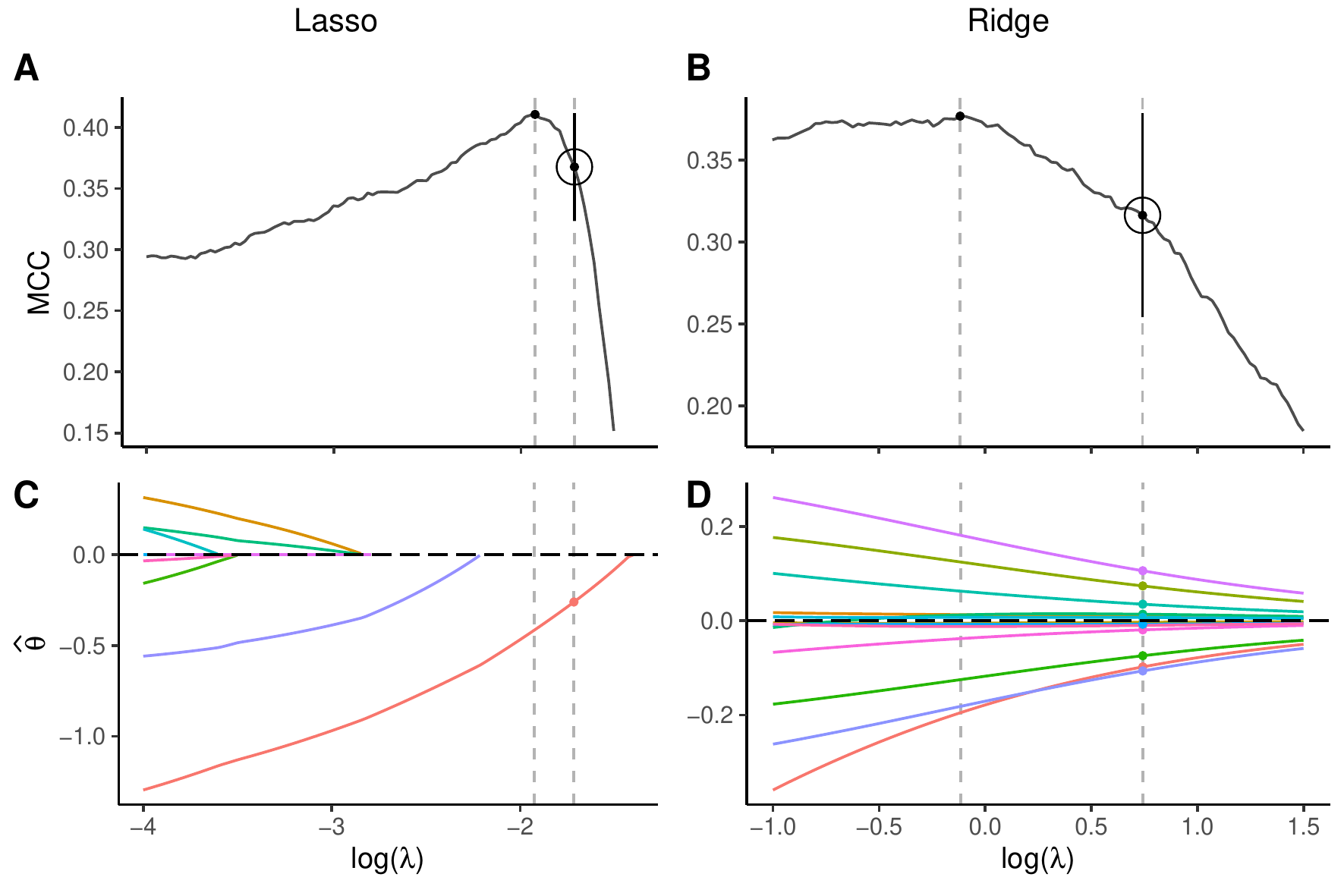}
	\caption{\label{fig:scat_reg} Penalised regression for scat classification models: lasso and ridge regression. Plots A and B show the cross-validated MCC estimates as a function of the logged regularisation parameter $\lambda$; the circled point is the largest $\lambda$ value (i.e., the most regularised model) within one standard error \eqref{eq:se_diff} of the value that maximises MCC. Plots C and D show the corresponding trajectories of the regularised estimates of the model parameters $\widehat{\theta}$, each denoted by a unique colour (labels ommited for clarity). The coloured points denote the final parameter estimates for the selected $\lambda$-values: for the ridge case, significant shrinkage is evident; for the lasso, just one non-zero predictor is retained ({\sf carbon-nitrogen ratio}).}
\end{figure}

As an alternative to selection amongst a discrete set of models, we illustrate the use of penalised regression \eqref{eq:elastic_net}, applying lasso ($\alpha = 1$) and ridge ($\alpha = 0$) regularisation to the global logistic model. In each case, we tune the regularisation parameter $\lambda$ using cross-validated MCC estimates with the same set of $K$-fold data splits used for the discrete approach. The CV-selected lasso model comprised just one (regularised) parameter associated with the predictor {\sf carbon-nitrogen ratio} (Figure \ref{fig:scat_reg}A). Despite its simplicity, the lasso model had a higher cross-validated MCC estimate than the ridge model which kept all 10 predictors (by construction), strongly regularising all of the associated parameters (Figure \ref{fig:scat_reg}B). We also applied elastic-net regularisation to the scat classification problem, however the optimal value of the CV-tuned hyperparameter $\alpha$ coincided with ordinary lasso case. We used the R package \texttt{glmnet} to fit the penalised regression models \citep{Friedman2010}.

\subsubsection*{Part 2: Regularising priors and Bayesian projective inference}

In a Bayesian setting, lasso-type regression can be implemented by an appropriate choice of prior distribution---to reflect the prior belief that only a proper subset of the included predictors are relevant. The direct analogue of the frequentist lasso is the Laplacian (two-sided exponential) prior \citep{Hastie2015}. An alternative choice is the regularised horseshoe prior \citep{Piironen2017hs}---this can be viewed as a continuous version of the traditional spike-and-slab approach (see \citet{Ohara2009} for background)---which provides support for a proper subset of parameters to be far from zero, using prior information to guide both the number and degree of regularisation of the non-zero parameter estimates. Using the \texttt{brms} package (see supplementary materials for details), we fit the global logistic scat model using: (1) Laplacian priors (i.e., LASSO), (2) regularised horseshoe priors, and (3) weakly-informative Gaussian priors (i.e., weak ridge-type regression). Based on PSIS-LOO estimates, the horseshoe variant was the best-performing model, followed by the lasso ($\Delta S = 2.1$, $\sigma^{\mathrm{diff}}=0.9$), and the weakly-informative priors ($\Delta S = 7.9$, $\sigma^{\mathrm{diff}}=3.4$). Figure S\ref{fig:scat_reg_post} shows the posterior distributions for all 11 parameters for each of the three models; the strongly regularising effect of the horseshoe prior is clearly visible, leaving only one variable ({\sf carbon-nitrogen ratio}) with support far from zero.

\begin{figure}[t]
	\centering
		\includegraphics[width=1\textwidth]{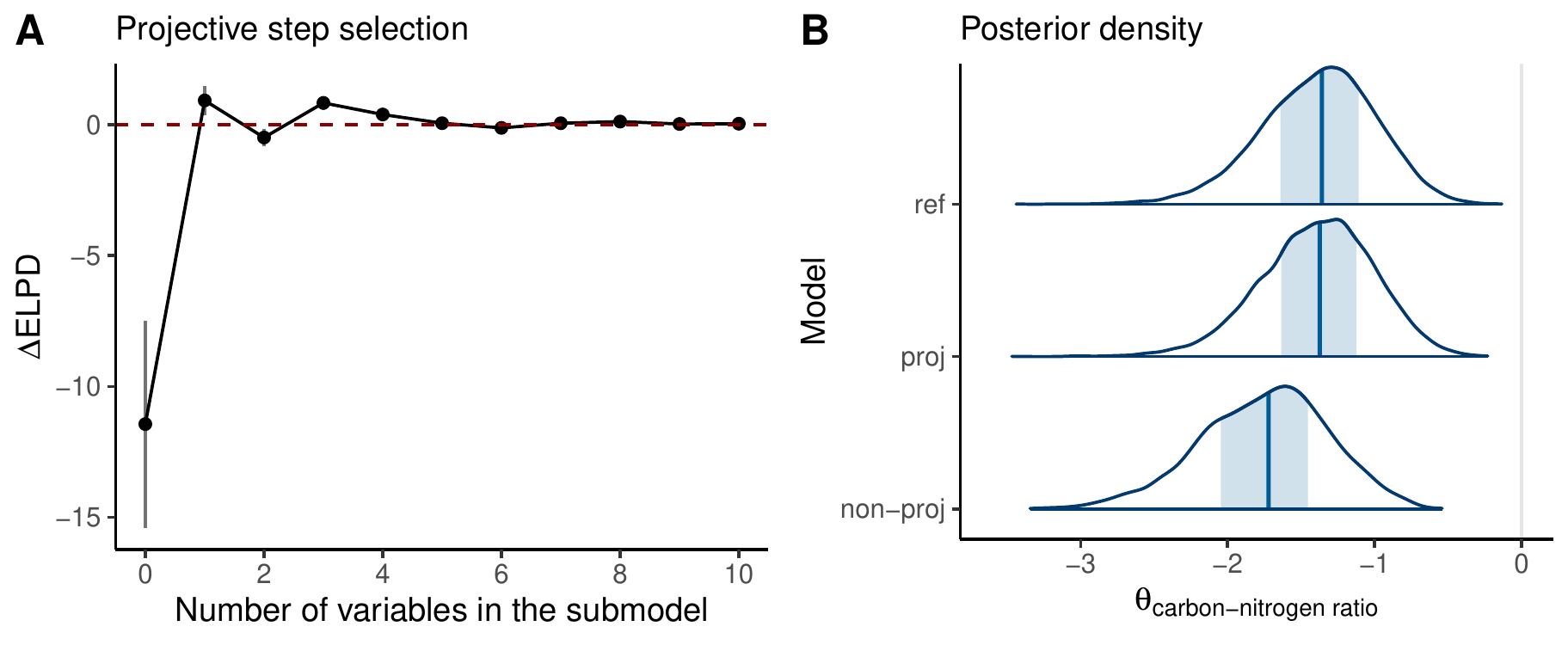}
	\caption{\label{fig:scat_bpi} Bayesian projective inference applied to Bernoulli scat-classification models. Plot A shows the mean and standard error of PSIS-LOO difference estimate $\Delta\mathrm{ELPD}$, with respect to the reference model (dashed line), for each increment of the projective step-selection process. The simplest submodel, with performance comparable to the reference model, contains only one-predictor: {\sf carbon-nitrogen ratio}. Plot B shows the marginal posterior distribution for the effect size of the {\sf carbon-nitrogen ratio} predictor for the reference model (ref) and the projected (proj) and non-projected (non-proj) one-parameter submodels.} 
\end{figure}

To assess the merits of using a simpler submodel, we apply BPI (see \emph{Regularisation, parsimony and all that} for details). Using the \texttt{projpred} package \citep{projpred2020} and taking as a reference model the fitted regularised horseshoe model, we use PSIS-LOO-based forward step selection to determine the optimal size of the alternative submodel and which variables to include. The selection results are shown in Figure \ref{fig:scat_bpi}A, confirming the intuition based on visual assessment, that the inclusion of just one variable ({\sf carbon-nitrogen ratio}) will have the best predictive performance (here even better than the reference model). The final step in BPI is to project the global posterior onto the selected submodel. Using the full set of posterior draws for the projection, Figure \ref{fig:scat_bpi}B shows a comparison of the marginal posterior densities of the {\sf carbon-nitrogen ratio} parameter for the reference model, the projected submodel, and the non-projected submodel (i.e., the same one-predictor submodel fit to the original data as if it were known a priori to be the preferred model). The reference and projected models have almost identical distributions, whereas the mean of non-projected distribution is appreciably further from zero than the other two. Although in other instances the reference and projected distributions may be less similar, the location and uncertainty of parameters estimates in the full model are generally good approximations to those of the corresponding parameters in a simpler submodel after correct propagation of model-selection uncertainty \citep{Dormann2018}, although clearly (see Figure S\ref{fig:scat_reg_post}) the degree of regularisation of the global model is an important consideration. Here we see the larger effect size of the non-projected model arising as a selection-induced bias, due to a failure to account for selection uncertainty. 

\subsubsection*{Part 3: nested CV to compare logistic and random forests models}
Machine-learning methods are increasingly used for classification problems, but their non-parametric (or `black-box') nature makes comparison with parametric approaches difficult. Here we illustrate the use of CV to compare the classification skill of a discretely selected parametric modelling approach (as per part 1) to a tuned machine-learning random forest algorithm (an ensemble of decision trees) \citep{Breiman2001}. We use nested CV (see Box 2) to completely separate model training, which includes variable selection and hyperparameter tuning, from the estimation of model performance. For the outer folds, we use 10-fold CV, repeated 50 times, and for the inner folds ordinary 10-fold CV. Thus, for each of the 500 outer training sets, 10 inner folds are used to select variables for the (discrete) logistic models, to tune the tree depth for the random forest models, and to tune the probability threshold for all models (see Box 1 - \emph{Tunable threshold}). Each repetition of the outer 10-fold CV generates a single MCC estimate for each selection approach. To reduce computation time, we set the number of trees in the random forest to the fixed value 500 and we included only the top 10\% of the 1024 discrete logistic models based on the MCC scores estimated in part 1. The analysis took 280 seconds using 40 Intel i7 cores.

The estimated mean MCC scores are $0.25$ and $0.375$ for the parametric and random-forest selection algorithms, respectively, and the modified standard error of their difference \eqref{eq:se_mod} is $0.06$. The performance difference between the two approaches is larger than the (modified) standard error estimate of the score difference, illustrating the trade-off between the interpretability of the parametric approach and the increased predictive skill of the machine-learning algorithm. Non-linear semi-parametric techniques, such as smoothing splines, could potentially increase the predictive performance of the logistic models, while visualisation techniques such as partial-dependence plots improve the interpretation of machine-learning approaches, thus softening the apparent trade-off between prediction and inference \citep{Efron2020}. 


\subsection*{Pinfish growth}

\textbf{Dataset}\\[2mm]
Name (R package): pinfish (fishmethods)\\
Description: Length, age and sex data for pinfish (\emph{Lagodon rhomboides}) from Tampa Bay, Florida.

\begin{figure}[t!]
	\centering
		\includegraphics[width=\textwidth]{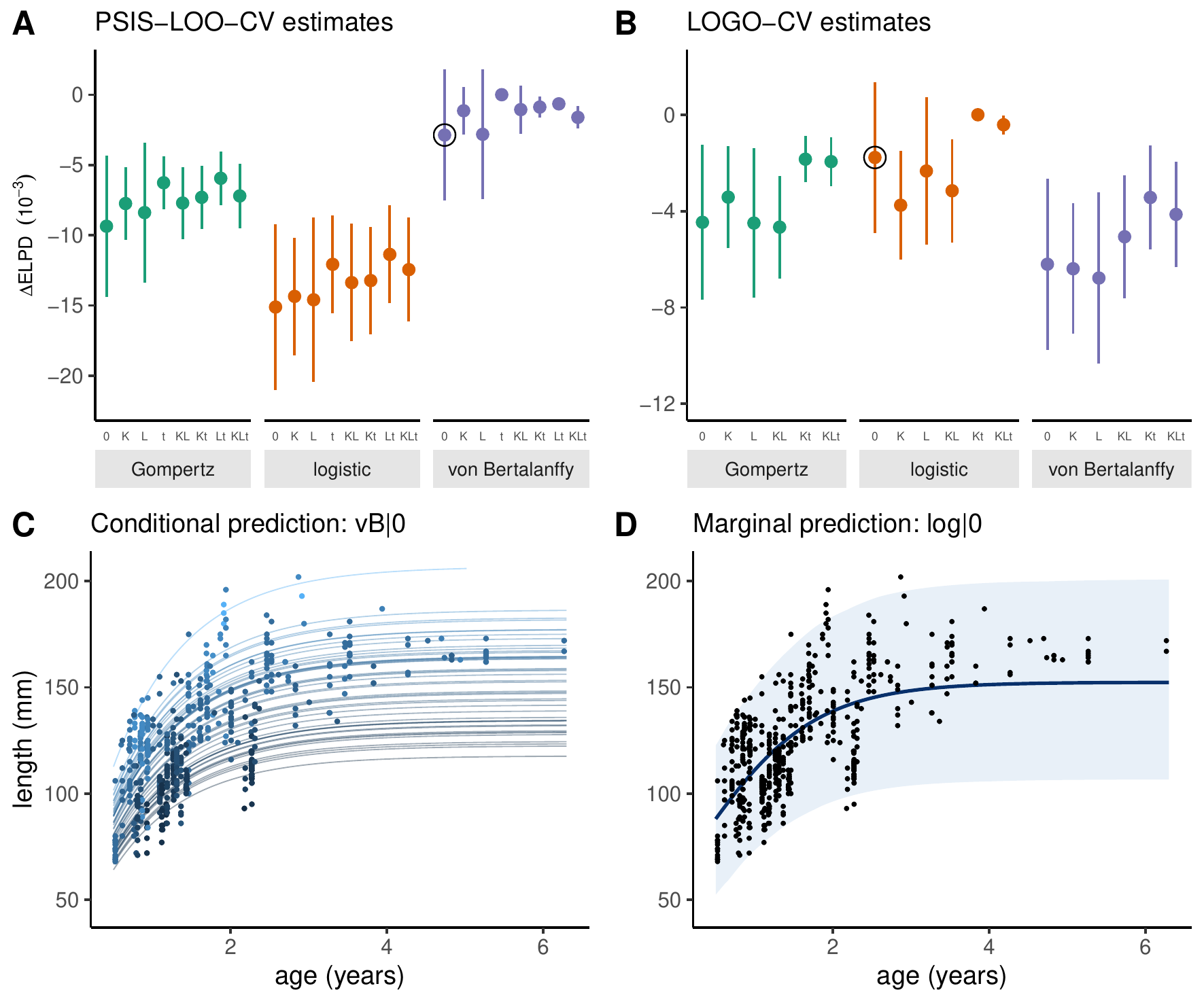}
	\caption{\label{fig:pinfish_panel} Model comparison and predicted curves for pinfish growth models. The plots on the left (A and C) and right (B and D) relate to a conditional and marginal focus, respectively. Applying the modified OSE rule to each focus, the fitted growth curves of the selected models (circled in A and B) are plotted alongside the data in C and D. The data and curves in C are coloured by haul according to the modelled haul-level length asymptote $L_{0,\,\mathrm{haul}}$. The dots and bars in A and C are the mean and standard error (\ref{eq:se_diff}) of the score differences, respectively. The model variants $\mbox{\textbf{x}}\mathbf{|t}$ and $\mbox{\textbf{x}}\mathbf{|Lt}$ are omitted from D for clarity due to their mean $\Delta\mathrm{ELPD}$ estimates being less than -20. The envelope in D is the 95\% credible interval. }
\end{figure}

\textbf{Analysis}\\[2mm]
The aim of the analysis is to determine which allometric growth function (see below) best describes the relationship between the available {\sf length} and {\sf age} measurements of pinfish, while accounting for the effect of {\sf sex} and {\sf haul} on the model parameters (the data comprise measurements from 45 separate fishery hauls). We take as candidate growth models the following commonly used non-linear functions \citep{Tjorve2010}:
\begin{equation}
    \begin{array}{ll}
    \mbox{Gompertz (G)} &L_{\mbox{\scriptsize{G}}}(a) = L_0 e^{-e^{(-K(a - t_0))}}\\
    \mbox{logistic (log)} &L_{\mbox{\scriptsize{log}}}(a) = {L_0}/(1 + e^{-K(a - t_0)})\\
    \mbox{von Bertalanffy (vB)} \qquad\qquad 
    &L_{\mbox{\scriptsize{vB}}}(a) = L_0(1 - e^{-K(a - t_0)}),
    \end{array}
\end{equation}
where $a$ is the age and $L$ is the (modelled) length. For each function, the parameters $L_0$, $K$, and $t_0$ denote the length asymptote, the growth rate, and the initial length, respectively. From inspection of Figure \ref{fig:pinfish_panel}C, there is an obvious effect of {\sf haul} on $L_0$; thus, we model $L_0$ hierarchically, using {\sf haul}-level intercepts (i.e., Gaussian distributed random effects). For each growth function $\textbf{x}= \mbox{G, log, vB}$, the most complex model $\mbox{\textbf{x}}\mathbf{|LKt}$ includes the two-level predictor {\sf sex} as a fixed effect on $L_0$, $K$, and $t_0$, specified as follows:\\[2mm]
\begin{equation}\label{eq:pinfish_bayesian_models}
	\mbox{\textbf{x}}\mathbf{|LKt}\qquad\qquad
    \begin{array}{cll}\\[-13mm]
	y_i &\sim& N(\mu_i,\sigma)\\[-2mm]
	\mu_i &=& L_{\mathrm{x}}(a_i;L_{0,\,i},K_{i},t_{0,\,i})\\[-2mm]
	L_{0,\,i} &=& \beta_{L,\,0} + \beta_{L,\,\mathrm{sex_1}[i]} + b_{\mathrm{haul}[i]}\\[-2mm]
	b &\sim& N(0,\tau)\\[-2mm]
	K_i &=& \beta_{K,0} + \beta_{K,\,\mathrm{sex_1}[i]}\\[-2mm]
	t_{0,\,i} &=&\beta_{t,0}  + \beta_{t,\,\mathrm{sex_1}[i]},
    \end{array}\hspace{3cm}
\end{equation}
from which simpler submodels are obtained by setting a subset of the terms $\beta_{\cdot,\,\mathrm{sex}_1}$ to zero. For example, $\mbox{\textbf{vB}}\mathbf{|K}$ is the von Bertalanffy growth model including sex as a fixed effect on $K$ only (i.e., $\beta_{L,\,\mathrm{sex}_1}=\beta_{t,\,\mathrm{sex}_1}=0$). All models include the {\sf haul}-level $L_0$ intercepts $b_{\mathrm{haul}[i]}$. We fit all 24 models in a Bayesian framework using R package \texttt{brms} \citep{brms}. Model priors were normal or student-$t$ distributions with standard deviations in the range of 0.3 to 10---sufficiently narrow to aid chain convergence while large enough to exert minimal influence on the posterior distributions (see supplementary materials for further details).

We illustrate two approaches to model selection using CV:\\[-10mm] 
\begin{enumerate}
\item conditional focus using approximate PSIS-LOO CV; and
\item marginal focus using leave-one-group-out (LOGO) CV.
\end{enumerate}

The conditional focus concerns model predictions to existing hauls, conditional on the {\sf haul}-level $L_0$ estimates which are treated as (regularised) model parameters. In this scenario, individual fish measurements constitute conditionally independent test samples, thereby permitting the use of LOO CV to estimate and compare model performance. A possible interpretation of this focus is that, after accounting for unmeasured effects of {\sf haul} on the data-generating process (e.g., changes in the measurement process due to differing apparatus or human expertise), estimates of conditional model performance quantify the capacity of candidate growth functions to explain within-haul growth trends and variation. 

The marginal focus concerns model predictions to new hauls, where the estimate of between-haul variation, $\tau$ in \eqref{eq:pinfish_bayesian_models}), is used in combination with the residual variation $\sigma$ to explain the total variability of the modelled response around the mean population-level growth curve. The use of LOGO CV in this case provides a direct means to estimate the predictive merits of the marginalised model, without having to actually integrate (i.e., marginalise) the likelihood over the hierarchical distribution of {\sf haul}-level intercepts (this can be slow and difficult for non-linear models). A possible interpretation of the marginal focus is that having estimated the {\sf haul}-level variability of the length asymptote, LOGO-CV quantifies the capacity of candidate growth functions to model the population-level mean growth curve, predicting to new fish measurements in new hauls.

The model-selection results for both types of focus are shown in Figures \ref{fig:pinfish_panel}A and \ref{fig:pinfish_panel}B, where the modified OSE rule has been applied using the standard error of the estimated score differences \eqref{eq:se_diff} which is implemented by default in \texttt{brms}. The best-performing growth model differed between the two types of focus, highlighting the importance of selecting a data-splitting scheme according to the predictive goals (e.g., the focus, in a random-effects context) of the analysis. However, in both types of focus, the application of the modified OSE rule suggested selection of the least complex variant $\mbox{\textbf{x}}\mathbf{|0}$ which excluded sex as a fixed effect on any of the model parameters. This does not mean that sex does not have an effect on fish growth, only that, given the available data and the specified set of candidate models, there is a risk of overfitting in selecting the more complex models that include sex dependence even if their estimated mean scores are slightly higher. We note that this risk is generally much lower in a Bayesian setting, in contrast to a frequentist approach, due to the regularising effect of both model priors and integration over the posterior distribution \citep{BDA3}.

Figures \ref{fig:pinfish_panel}C and \ref{fig:pinfish_panel}D show the predicted curves for each focus superposed on the data. Viewed as posterior predictive checks, these plots illustrate the capacity of the models to generate plausible data. For example, the modelled asymptote in Figure \ref{fig:pinfish_panel}D is lower than all of the data points for fish older than four years, suggesting that model fitting and model-performance estimates have been influenced by the predominance of data for younger fish. The 95\% credible interval appears to provide nominal coverage of the data in the marginal prediction.

In many instances, researchers will have an \textit{a priori} reason to model a group-level effect as either fixed or random. For example, random effects may be chosen to `borrow strength' across group levels, to account for unmeasured group-level effects that vary around a population-level mean, to partition variance via marginalisation, or to provide regularisation to improve model prediction and/or convergence of the parameter-estimation algorithm \citep{Hobbs2015}. It is possible, however, to use the data (via CV) to investigate whether a random effect or a fixed effect is the preferred choice, or at least to compare the two sets of estimates (e.g., to determine the amount of shrinkage). To illustrate, we apply PSIS-LOO to the fixed and random variants of the conditional model $\mbox{\textbf{vB}}\mathbf{|0}$, obtaining an estimated score difference of $2.17$ ($\sigma_{\mathrm{diff}} = 1.85$) in favour of the random model. Using \eqref{eq:p_loo}, the estimated effective number of parameters $p_{\mathrm{loo}}$ for the random model is $42.6$ compared to $45.7$ for the fixed model---the latter is less than the nominal count of 48 fixed parameters due to the inherent regularisation of the Bayesian estimation process. The weak preference for the random model, and the reduction of just $3.1$ (SE = $0.6$) effective parameters, is reflected in the mild shrinkage of the haul-level $L_0$ estimates for the random model compared to the fixed (Figure \ref{fig:shrinkage}).   

For clarity of exposition, we have omitted more complex models, such as those including haul-level intercepts for $K_0$ and $t_0$. The inclusion of more than one group-level effect will generally lead to a multivariate hierarchical structure with corresponding correlation terms. However, assuming there are sufficient data to estimate the model parameters, model selection using CV can be implemented in these more complex cases in the same way as we have demonstrated above.

\section*{Discussion}

We have presented a comprehensive primer on understanding and applying cross-validation for model selection, with a focus on how ecologists would use this approach. This includes an overview of commonly used techniques, theoretical aspects and recent developments, as well as practical guidance for implementation. Although it is difficult to provide universal advice for using CV, since the choice of data-splitting scheme and score depends on the modelling context, in most instances we recommend leave-one-out or approximate LOO to minimise bias. Otherwise we recommend $K$-fold with $K$ set as large as practicably possible. If $K<10$ bias correction should be used when available. Further, the use of blocked, LOGO, or stratified data splitting should be investigated when the data are autocorrelated or hierarchically structured. To mitigate overfitting, we recommend calibrated selection via the modified OSE rule \citep{Yates2021} which accounts for the predominant cause of overfitting: score-estimation uncertainty. 

There are many software packages for the R programming language that aid the implementation of CV. The packages \texttt{caret} \citep{caret} and \texttt{tidymodels} \citep{tidymodels} are particularly versatile, incorporating a broad class of model types (via commonly used auxiliary packages) ranging from generalised linear models and penalised regression to (tuned) machine-learning methods and certain classes of Bayesian models. The \texttt{tidymodels} package is a composite of several helper packages including \texttt{rsample} for data splitting (e.g., nested $K$-fold) and \texttt{yardstick} for estimating both pre- and user-defined model scores; these packages can be used independently of the integrated workflow of the parent package. Bias-corrected $K$-fold CV can be performed using the packages \texttt{bestglm} \citep{bestglm} or \texttt{boot} \citep{boot}, although support is limited to generalised linear models; the latter permits the use of custom loss functions. For fully Bayesian approaches, the \texttt{loo} \citep{Vehtari2017} package implements the approximate method PSIS-LOO, requiring as an input the log-likelihood evaluated at a set of posterior simulations of the parameter estimates. All of the aforementioned packages support parallel processing. When existing software is unsuitable, custom code can be developed. Generally speaking, if the set of candidate models is able to be fit to a training subset of the data and subsequently predicted to a test set, then CV implementation is usually straight-forward, requiring only data-splitting, repeated model-fitting, and subsequent aggregation of the score estimates. Existing software can help with each of these steps. 

The ever-increasing uptake of Bayesian methodology in ecology in recent decades has been facilitated by the steady publication of both pedagogical texts (e.g., \citet{Hoeting1999,Ellison2004,BDA3,Hobbs2015}) and integrated software (e.g., BUGS, JAGS, PyMC, Stan). Recently, there has been a surge in the development of Bayesian model-selection techniques (e.g., \citet{Vehtari2017, Burkner2020, Burkner2021}), wherein IT methods based on CV have played a central role. The parallel development of Hamiltonian Monte Carlo \citep[HMC,][]{Neal2011} sampling methods has brought significant efficiency gains which are readily accessible to ecologists via the \texttt{rstan} \citep{rstan} package or its user-friendly front-end packages \texttt{rstanarm} \citep{rstanarm} and \texttt{brms} \citep{brms}. The combination of HMC with PSIS-LOO permits highly efficient model fitting and score estimation for a broad range of model-selection contexts, including historically challenging model classes such as non-factorised data models (e.g., observation-level latent-variable models)\citep{Burkner2021}---equivalent methods do not exist outside of the Bayesian setting. None of these implementations require informative priors, although they could certainly be used where appropriate. Thus, there is a compelling case for the use of Bayesian methods for model specification and model selection in ecology (at least for likelihood-based inference) where the increasing use of complex hierarchical model structures is a natural fit with sampling-based estimation which has robust estimation properties and a high tolerance for model complexity.

In most model-selection problems, CV can be used to estimate scores and suggest a preferred model, but is the selected model good enough and can we make valid inference after selection? In terms of model checking, it is important to note that CV is not a replacement for data simulation (e.g., posterior predictive checks or the parametric bootstrap) or within-sample measures such as the proportion of deviance explained or graphical checks of residual distributions. These techniques are used to assess the adequacy of the selected model to generate plausible data where expert knowledge is often required to determine what constitutes adequate in a giving modelling context \citep{Gabry2019}. Given the selected model has passed model checking, the issue of valid inference remains. Although it is common practice, it is generally problematic to make inferences using the parameter estimates of the selected model fit to the full data set without accounting in some way for model-selection uncertainty---problems include inflated effect sizes (i.e., selection-induced bias) and underestimated uncertainty intervals \citep{Hjort2003}. BPI (see \emph{Regularisation, parsimony and all that} and \emph{Examples}) provides a practical solution to the problem of valid inference and some recently developed methods can be applied within other model-selection settings \citep{Claeskens2008, Berk2013, Lee2016, Charkhi2018}; see \citet{Yates2021} for further discussion.

Cross-validation is clearly a practical and versatile technique for model selection in ecology: a research field in which statistical inference has been increasingly dominated over the last two decades by the use of information criteria (especially AIC). Yet given the broad applicability of CV, and the now near-ubiquitous availability of multi-core parallel computing, one might ask whether the prominent role of IC is coming to an end. Information Criteria are generally easier and faster to compute than conventional CV estimates (except LOO in linear regression), but the need for bias correction imposes limitations on the data and models, leading to an ever-growing set of IC variants to accommodate specialised settings; e.g., AICc, mAIC, AICq, QAIC, TIC, WAIC, or indeed just run through the alphabet: AIC (Akaike), BIC ([pseudo-]Bayesian), CIC (Covariance), DIC (Deviance), EIC (Empirical), FIC (Focused), etc. High-performance parallel computation and the development of accurate approximate CV methods now makes CV comparable with IC for speed, obviating the need for such a large number of specialised variants, while at the same time increasing the reach of model-selection techniques into complex modelling contexts where traditional IC remain inapplicable (e.g., deep-learning approaches such as artificial neural networks). In a Bayesian setting, \citet{Vehtari2017} advocate for the use of (approximate) LOO over the widely applicable information criterion (WAIC) due to improved stability, although both are considered superior to the commonly used AIC and DIC. Given the demonstrable utility and generality of CV for comparing a diverse range of statistical models, fitted to simple or highly complex data sets, we foresee its increasingly widespread adoption in the domain of ecology.

\section*{Acknowledgements}
This work was funded by the Australian Research Council grant FL160100101. The authors would like to thank Leon Barmuta for helpful discussions. 

\section*{Code availability}
The code and online tutorial to reproduce the analyses in this manuscript are available at \url{https://l-a-yates.github.io/CVPrimer/}


\begin{thebibliography}{72}
\expandafter\ifx\csname natexlab\endcsname\relax\def\natexlab#1{#1}\fi
\expandafter\ifx\csname url\endcsname\relax
  \def\url#1{{\tt #1}}\fi
\expandafter\ifx\csname urlprefix\endcsname\relax\def\urlprefix{URL }\fi

\bibitem[{Afrabandpey et~al.(2020)Afrabandpey, Peltola, Piironen, Vehtari, and
  Kaski}]{Afrabandpey2020}
Afrabandpey, H., T.~Peltola, J.~Piironen, A.~Vehtari, and S.~Kaski.
\newblock 2020.
\newblock {A decision-theoretic approach for model interpretability in Bayesian
  framework}.
\newblock Machine Learning {\bf 109}:1855--1876.

\bibitem[{Aho et~al.(2014)Aho, Derryberry, and Peterson}]{aho2014}
Aho, K., D.~Derryberry, and T.~Peterson.
\newblock 2014.
\newblock Model selection for ecologists: the worldviews of AIC and BIC.
\newblock Ecology {\bf 95}:631--636.

\bibitem[{Akaike(1973)}]{Akaike1973}
Akaike, H., 1973.
\newblock {Information theory and an extension of the maximum likelihood
  principle}.
\newblock Pages 267--281 {\em in\/} {B. N. Petrov and F. Csaki}, editor. Second
  International Symposium on Information Theory (Tsahkadsor, 1971). Academiai
  Kiado, Budapest.

\bibitem[{Allouche et~al.(2006)Allouche, Tsoar, and Kadmon}]{Allouche2006}
Allouche, O., A.~Tsoar, and R.~Kadmon.
\newblock 2006.
\newblock {Assessing the accuracy of species distribution models: prevalence,
  kappa and the true skill statistic (TSS)}.
\newblock Journal of Applied Ecology {\bf 43}:1223--1232.

\bibitem[{Arlot and Celisse(2010)}]{Arlot2010}
Arlot, S., and A.~Celisse.
\newblock 2010.
\newblock {A survey of cross-validation procedures for model selection}.
\newblock Statistics Surveys {\bf 4}:40--79.

\bibitem[{Arlot and Lerasle(2016)}]{Arlot2016}
Arlot, S., and M.~Lerasle.
\newblock 2016.
\newblock {Choice of V for V-fold cross-validation in least-squares density
  estimation}.
\newblock Journal of Machine Learning Research {\bf 17}:1--50.

\bibitem[{Berk et~al.(2013)Berk, Brown, Buja, Zhang, and Zhao}]{Berk2013}
Berk, R., L.~Brown, A.~Buja, K.~Zhang, and L.~Zhao.
\newblock 2013.
\newblock {Valid post-selection inference}.
\newblock Annals of Statistics {\bf 41}:802--837.

\bibitem[{Bernardo(1979)}]{Bernardo1979}
Bernardo, J.~M.
\newblock 1979.
\newblock Expected Information as Expected Utility.
\newblock The Annals of Statistics {\bf 7}.

\bibitem[{Bolker et~al.(2009)Bolker, Brooks, Clark, Geange, Poulsen, Stevens,
  and White}]{Bolker2009}
Bolker, B.~M., M.~E. Brooks, C.~J. Clark, S.~W. Geange, J.~R. Poulsen, M.~H.~H.
  Stevens, and J.~S.~S. White, 2009.
\newblock {Generalized linear mixed models: a practical guide for ecology and
  evolution}.

\bibitem[{Breiman(2001)}]{Breiman2001}
Breiman, L.
\newblock 2001.
\newblock Machine Learning {\bf 45}:5--32.

\bibitem[{Breiman et~al.(1984)Breiman, Friedman, Olshen, and
  Stone}]{Breiman1984}
Breiman, L., J.~H. Friedman, R.~A. Olshen, and C.~J. Stone.
\newblock 1984.
\newblock {Classification and regression trees}.
\newblock Wadsworth {\&} Brooks/Cole Advanced Books {\&} Software,.

\bibitem[{Breiman and Spector(1992)}]{Breiman1992}
Breiman, L., and P.~Spector.
\newblock 1992.
\newblock {Submodel Selection and Evaluation in Regression. The X-Random Case}.
\newblock International Statistical Review / Revue Internationale de
  Statistique {\bf 60}:291.

\bibitem[{Bürkner(2017)}]{brms}
Bürkner, P.-C.
\newblock 2017.
\newblock {brms}: An {R} Package for {Bayesian} Multilevel Models Using {Stan}.
\newblock Journal of Statistical Software {\bf 80}:1--28.

\bibitem[{B{\"{u}}rkner et~al.(2020)B{\"{u}}rkner, Gabry, and
  Vehtari}]{Burkner2020}
B{\"{u}}rkner, P.-C., J.~Gabry, and A.~Vehtari.
\newblock 2020.
\newblock {Approximate leave-future-out cross-validation for Bayesian time
  series models}.
\newblock Journal of Statistical Computation and Simulation {\bf
  90}:2499--2523.

\bibitem[{B{\"{u}}rkner et~al.(2021)B{\"{u}}rkner, Gabry, and
  Vehtari}]{Burkner2021}
B{\"{u}}rkner, P.~C., J.~Gabry, and A.~Vehtari.
\newblock 2021.
\newblock {Efficient leave-one-out cross-validation for Bayesian non-factorized
  normal and Student-t models}.
\newblock Computational Statistics {\bf 36}:1243--1261.

\bibitem[{Burman(1989)}]{Burman1989}
Burman, P.
\newblock 1989.
\newblock {A comparative study of ordinary cross-validation, v-fold
  cross-validation and the repeated learning-testing methods}.
\newblock Biometrika {\bf 76}:503--514.

\bibitem[{Burman et~al.(1994)Burman, Chow, and Nolan}]{Burman1994}
Burman, P., E.~Chow, and D.~Nolan.
\newblock 1994.
\newblock {A cross-validatory method for dependent data}.
\newblock Biometrika {\bf 81}:351--358.

\bibitem[{Burnham and Anderson(2002)}]{Burnham2002}
Burnham, K.~P., and D.~R. Anderson.
\newblock 2002.
\newblock {Model selection and multimodel inference: a practical
  information-theoretic approach}.
\newblock Second edition.
\newblock Springer-Verlag, New York.

\bibitem[{Canty and Ripley(2021)}]{boot}
Canty, A., and B.~D. Ripley, 2021.
\newblock boot: Bootstrap R (S-Plus) Functions.

\bibitem[{Cawley and Talbot(2010)}]{Cawley2010}
Cawley, G.~C., and N.~L. Talbot.
\newblock 2010.
\newblock {On over-fitting in model selection and subsequent selection bias in
  performance evaluation}.
\newblock Journal of Machine Learning Research {\bf 11}:2079--2107.

\bibitem[{Charkhi and Claeskens(2018)}]{Charkhi2018}
Charkhi, A., and G.~Claeskens.
\newblock 2018.
\newblock {Asymptotic post-selection inference for the Akaike information
  criterion}.
\newblock Biometrika {\bf 105}:645--664.

\bibitem[{Chicco et~al.(2021)Chicco, Warrens, and Jurman}]{Chicco2021}
Chicco, D., M.~J. Warrens, and G.~Jurman.
\newblock 2021.
\newblock {The Matthews Correlation Coefficient (MCC) is More Informative Than
  Cohen's Kappa and Brier Score in Binary Classification Assessment}.
\newblock IEEE Access {\bf 9}:78368--78381.

\bibitem[{Claeskens and Hjort(2008)}]{Claeskens2008}
Claeskens, G., and N.~L. Hjort.
\newblock 2008.
\newblock {Model Selection and Model Averaging}.
\newblock Cambridge Series in Statistical and Probabilistic Mathematics,
  Cambridge University Press.

\bibitem[{Davison and Hinkley(1997)}]{Davison1997}
Davison, A.~C., and D.~V. Hinkley.
\newblock 1997.
\newblock {Bootstrap methods and their application}.
\newblock Cambridge University Press.

\bibitem[{Deng(2012)}]{Deng2012}
Deng, Y.
\newblock 2012.
\newblock Applied Parallel Computing.
\newblock World Scientific Publishing, Singapore, Singapore.

\bibitem[{Dormann et~al.(2018)Dormann, Calabrese, Guillera-Arroita, Matechou,
  Bahn, Barto{\'{n}}, Beale, Ciuti, Elith, Gerstner, Guelat, Keil,
  Lahoz-Monfort, Pollock, Reineking, Roberts, Schr{\"{o}}der, Thuiller, Warton,
  Wintle, Wood, W{\"{u}}est, and Hartig}]{Dormann2018}
Dormann, C.~F., J.~M. Calabrese, G.~Guillera-Arroita, E.~Matechou, V.~Bahn,
  K.~Barto{\'{n}}, C.~M. Beale, S.~Ciuti, J.~Elith, K.~Gerstner, J.~Guelat,
  P.~Keil, J.~J. Lahoz-Monfort, L.~J. Pollock, B.~Reineking, D.~R. Roberts,
  B.~Schr{\"{o}}der, W.~Thuiller, D.~I. Warton, B.~A. Wintle, S.~N. Wood, R.~O.
  W{\"{u}}est, and F.~Hartig.
\newblock 2018.
\newblock {Model averaging in ecology: a review of Bayesian,
  information-theoretic, and tactical approaches for predictive inference}.
\newblock Ecological Monographs {\bf 88}:485--504.

\bibitem[{Dormann et~al.(2013)Dormann, Elith, Bacher, Buchmann, Carl, Carré,
  Marquéz, Gruber, Lafourcade, Leitão, Münkemüller, McClean, Osborne,
  Reineking, Schröder, Skidmore, Zurell, and Lautenbach}]{Dormann2013}
Dormann, C.~F., J.~Elith, S.~Bacher, C.~Buchmann, G.~Carl, G.~Carré, J.~R.~G.
  Marquéz, B.~Gruber, B.~Lafourcade, P.~J. Leitão, T.~Münkemüller,
  C.~McClean, P.~E. Osborne, B.~Reineking, B.~Schröder, A.~K. Skidmore,
  D.~Zurell, and S.~Lautenbach.
\newblock 2013.
\newblock Collinearity: a review of methods to deal with it and a simulation
  study evaluating their performance.
\newblock Ecography {\bf 36}:27--46.

\bibitem[{Efron(2020)}]{Efron2020}
Efron, B.
\newblock 2020.
\newblock {Prediction, Estimation, and Attribution}.
\newblock Journal of the American Statistical Association {\bf 115}:636--655.

\bibitem[{Ellison(2004)}]{Ellison2004}
Ellison, A.~M.
\newblock 2004.
\newblock Bayesian inference in ecology.
\newblock Ecology Letters {\bf 7}:509--520.

\bibitem[{Fang(2011)}]{Fang2011}
Fang, Y.
\newblock 2011.
\newblock {Asymptotic Equivalence between Cross-Validations and Akaike
  Information Criteria in Mixed-Effects Models}.
\newblock Journal of Data Science {\bf 9}:15--21.

\bibitem[{Fletcher and Fortin(2018)}]{Fletcher2018}
Fletcher, R., and M.-J. Fortin.
\newblock 2018.
\newblock Spatial Ecology and Conservation Modeling.
\newblock Springer International Publishing.

\bibitem[{Friedman et~al.(2010)Friedman, Hastie, and Tibshirani}]{Friedman2010}
Friedman, J., T.~Hastie, and R.~Tibshirani.
\newblock 2010.
\newblock Regularization Paths for Generalized Linear Models via Coordinate
  Descent.
\newblock Journal of Statistical Software {\bf 33}:1--22.

\bibitem[{Gabry et~al.(2019)Gabry, Simpson, Vehtari, Betancourt, and
  Gelman}]{Gabry2019}
Gabry, J., D.~Simpson, A.~Vehtari, M.~Betancourt, and A.~Gelman.
\newblock 2019.
\newblock Visualization in Bayesian workflow.
\newblock Journal of the Royal Statistical Society: Series A (Statistics in
  Society) {\bf 182}:389--402.

\bibitem[{Garthwaite and Mubwandarikwa(2010)}]{Garthwaite2010}
Garthwaite, P.~H., and E.~Mubwandarikwa.
\newblock 2010.
\newblock {Selection of weights for weighted model averaging}.
\newblock Australian and New Zealand Journal of Statistics {\bf 52}:363--382.

\bibitem[{Gelman et~al.(2013)Gelman, Carlin, Stern, Dunson, Vehtari, and
  Rubin}]{BDA3}
Gelman, A., J.~B. Carlin, H.~S. Stern, D.~B. Dunson, A.~Vehtari, and D.~B.
  Rubin.
\newblock 2013.
\newblock Bayesian Data Analysis.
\newblock Hardcover edition.
\newblock Chapman and Hall/CRC.

\bibitem[{Gneiting(2011)}]{Gneiting2011}
Gneiting, T.
\newblock 2011.
\newblock {Making and evaluating point forecasts}.
\newblock Journal of the American Statistical Association {\bf 106}:746--762.

\bibitem[{Gneiting and Raftery(2007)}]{Gneiting2007}
Gneiting, T., and A.~E. Raftery.
\newblock 2007.
\newblock {Strictly proper scoring rules, prediction, and estimation}.
\newblock Journal of the American Statistical Association {\bf 102}:359--378.

\bibitem[{Goodrich et~al.(2020)Goodrich, Gabry, Ali, and Brilleman}]{rstanarm}
Goodrich, B., J.~Gabry, I.~Ali, and S.~Brilleman, 2020.
\newblock rstanarm: {Bayesian} applied regression modeling via {Stan}.

\bibitem[{Hastie et~al.(2009)Hastie, Tibshirani, and Friedman}]{Hastie2009}
Hastie, T., R.~Tibshirani, and J.~Friedman.
\newblock 2009.
\newblock {The Elements of Statistical Learning Data Mining, Inference, and
  Prediction}.
\newblock Second edition.
\newblock Springer Series in Statistics.

\bibitem[{Hastie et~al.(2015)Hastie, Tibshirani, and Wainwright}]{Hastie2015}
Hastie, T., R.~Tibshirani, and M.~Wainwright.
\newblock 2015.
\newblock Statistical Learning with Sparsity.
\newblock Chapman and Hall/{CRC}, {Boca Raton}.

\bibitem[{Hjort and Claeskens(2003)}]{Hjort2003}
Hjort, N.~L., and G.~Claeskens.
\newblock 2003.
\newblock {Frequentist Model Average Estimators}.
\newblock Journal of the American Statistical Association {\bf 98}:879--899.

\bibitem[{Hobbs and Hooten(2015)}]{Hobbs2015}
Hobbs, N.~T., and M.~B. Hooten.
\newblock 2015.
\newblock Bayesian models: a statistical primer for ecologists.
\newblock Princeton University Press.

\bibitem[{Hoeting et~al.(1999)Hoeting, Madigan, Raftery, and
  Volinsky}]{Hoeting1999}
Hoeting, J.~A., D.~Madigan, A.~E. Raftery, and C.~T. Volinsky.
\newblock 1999.
\newblock {Bayesian model averaging: A tutorial}.
\newblock Statistical Science {\bf 14}:382--401.

\bibitem[{Hooten et~al.(2015)Hooten, Hobbs, and Ellison}]{Hooten2015}
Hooten, M.~B., N.~T. Hobbs, and A.~M. Ellison.
\newblock 2015.
\newblock {A guide to Bayesian model selection for ecologists}.
\newblock Ecological Monographs {\bf 85}:3--28.

\bibitem[{James et~al.(2013)James, Witten, Hastie, and Tibshirani}]{James2013}
James, G., D.~Witten, T.~Hastie, and R.~Tibshirani.
\newblock 2013.
\newblock {An Introduction to Statistical Learning}.
\newblock Springer New York, New York, NY.

\bibitem[{Johnson(1999)}]{Johnson1999}
Johnson, D.~H.
\newblock 1999.
\newblock The Insignificance of Statistical Significance Testing.
\newblock The Journal of Wildlife Management {\bf 63}:763--772.

\bibitem[{Kuhn(2021)}]{caret}
Kuhn, M., 2021.
\newblock caret: Classification and Regression Training.

\bibitem[{Kuhn and Wickham(2020)}]{tidymodels}
Kuhn, M., and H.~Wickham, 2020.
\newblock Tidymodels: a collection of packages for modeling and machine
  learning using tidyverse principles.

\bibitem[{Lee et~al.(2016)Lee, Sun, Sun, and Taylor}]{Lee2016}
Lee, J.~D., D.~L. Sun, Y.~Sun, and J.~E. Taylor.
\newblock 2016.
\newblock {Exact post-selection inference, with application to the lasso}.
\newblock Annals of Statistics {\bf 44}:907--927.

\bibitem[{Luque et~al.(2019)Luque, Carrasco, Martín, and {de las
  Heras}}]{Luque2019}
Luque, A., A.~Carrasco, A.~Martín, and A.~{de las Heras}.
\newblock 2019.
\newblock The impact of class imbalance in classification performance metrics
  based on the binary confusion matrix.
\newblock Pattern Recognition {\bf 91}:216--231.

\bibitem[{Matthews(1975)}]{Matthews1975}
Matthews, B.
\newblock 1975.
\newblock Comparison of the predicted and observed secondary structure of T4
  phage lysozyme.
\newblock Biochimica et Biophysica Acta (BBA) - Protein Structure {\bf
  405}:442--451.

\bibitem[{McLeod and Xu(2020)}]{bestglm}
McLeod, A., and C.~Xu, 2020.
\newblock bestglm: Best Subset GLM.

\bibitem[{Merkle et~al.(2019)Merkle, Furr, and Rabe-Hesketh}]{Merkle2019}
Merkle, E.~C., D.~Furr, and S.~Rabe-Hesketh.
\newblock 2019.
\newblock Bayesian Comparison of Latent Variable Models: Conditional Versus
  Marginal Likelihoods.
\newblock Psychometrika {\bf 84}:802--829.

\bibitem[{Neal(2011)}]{Neal2011}
Neal, R., 2011.
\newblock {MCMC} Using Hamiltonian Dynamics.
\newblock  {\em in\/} Chapman {\&} Hall/{CRC} Handbooks of Modern Statistical
  Methods. Chapman and Hall/{CRC}.

\bibitem[{O'Hara and Sillanp{\"{a}}{\"{a}}(2009)}]{Ohara2009}
O'Hara, R.~B., and M.~J. Sillanp{\"{a}}{\"{a}}.
\newblock 2009.
\newblock {A review of bayesian variable selection methods: What, how and
  which}.
\newblock Bayesian Analysis {\bf 4}:85--118.

\bibitem[{Piironen et~al.(2020{\natexlab{{\em a\/}}})Piironen, Paasiniemi,
  Catalina, and Vehtari}]{projpred2020}
Piironen, J., M.~Paasiniemi, A.~Catalina, and A.~Vehtari, 2020{\natexlab{{\em
  a\/}}}.
\newblock projpred: Projection Predictive Feature Selection.

\bibitem[{Piironen et~al.(2020{\natexlab{{\em b\/}}})Piironen, Paasiniemi, and
  Vehtari}]{Piironen2020}
Piironen, J., M.~Paasiniemi, and A.~Vehtari.
\newblock 2020{\natexlab{{\em b\/}}}.
\newblock {Projective inference in high-dimensional problems: Prediction and
  feature selection}.
\newblock Electronic Journal of Statistics {\bf 14}:2155--2197.

\bibitem[{Piironen and Vehtari(2017{\natexlab{{\em a\/}}})}]{piironen2017b}
Piironen, J., and A.~Vehtari, 2017{\natexlab{{\em a\/}}}.
\newblock {On the Hyperprior Choice for the Global Shrinkage Parameter in the
  Horseshoe Prior}.
\newblock Pages 905--913 {\em in\/} A.~Singh and J.~Zhu, editors. Proceedings
  of the 20th International Conference on Artificial Intelligence and
  Statistics, volume~54 of {\em Proceedings of Machine Learning Research\/}.
  PMLR.

\bibitem[{Piironen and Vehtari(2017{\natexlab{{\em b\/}}})}]{Piironen2017hs}
Piironen, J., and A.~Vehtari.
\newblock 2017{\natexlab{{\em b\/}}}.
\newblock {Sparsity information and regularization in the horseshoe and other
  shrinkage priors}.
\newblock Electronic Journal of Statistics {\bf 11}:5018 -- 5051.

\bibitem[{Reid(2015)}]{Reid2015}
Reid, R. E.~B.
\newblock 2015.
\newblock A morphometric modeling approach to distinguishing among bobcat,
  coyote and gray fox scats.
\newblock Wildlife Biology {\bf 21}:254--262.

\bibitem[{Richards(2005)}]{Richards2005}
Richards, S.~A.
\newblock 2005.
\newblock {Testing ecological theory using the information-theoretic approach:
  Examples and cautionary results}.
\newblock Ecology {\bf 86}:2805--2814.

\bibitem[{Roberts et~al.(2017)Roberts, Bahn, Ciuti, Boyce, Elith,
  Guillera-Arroita, Hauenstein, Lahoz-Monfort, Schr{\"{o}}der, Thuiller,
  Warton, Wintle, Hartig, and Dormann}]{Roberts2017}
Roberts, D.~R., V.~Bahn, S.~Ciuti, M.~S. Boyce, J.~Elith, G.~Guillera-Arroita,
  S.~Hauenstein, J.~J. Lahoz-Monfort, B.~Schr{\"{o}}der, W.~Thuiller, D.~I.
  Warton, B.~A. Wintle, F.~Hartig, and C.~F. Dormann.
\newblock 2017.
\newblock {Cross-validation strategies for data with temporal, spatial,
  hierarchical, or phylogenetic structure}.
\newblock Ecography {\bf 40}:913--929.

\bibitem[{Shao(1993)}]{Shao1993}
Shao, J.
\newblock 1993.
\newblock {Linear model selection by cross-validation}.
\newblock Journal of Statistical Planning and Inference {\bf 128}:231--240.

\bibitem[{Sivula et~al.(2020)Sivula, Magnusson, and Vehtari}]{sivula2020}
Sivula, T., M.~Magnusson, and A.~Vehtari, 2020.
\newblock Uncertainty in Bayesian Leave-One-Out Cross-Validation Based Model
  Comparison.

\bibitem[{{Stan Development Team}(2020)}]{rstan}
{Stan Development Team}, 2020.
\newblock {RStan}: the {R} interface to {Stan}.

\bibitem[{Stone(1977)}]{Stone1977}
Stone, M.
\newblock 1977.
\newblock {An Asymptotic Equivalence of Choice of Model by Cross-Validation and
  Akaike's Criterion}.
\newblock Journal of the Royal Statistical Society: Series B (Methodological)
  {\bf 39}:44--47.

\bibitem[{Tj{\o}rve and Tj{\o}rve(2010)}]{Tjorve2010}
Tj{\o}rve, E., and K.~M. Tj{\o}rve.
\newblock 2010.
\newblock {A unified approach to the Richards-model family for use in growth
  analyses: Why we need only two model forms}.
\newblock Journal of Theoretical Biology {\bf 267}:417--425.

\bibitem[{Tredennick et~al.(2021)Tredennick, Hooker, Ellner, and
  Adler}]{Tredennick2021}
Tredennick, A.~T., G.~Hooker, S.~P. Ellner, and P.~B. Adler.
\newblock 2021.
\newblock A practical guide to selecting models for exploration, inference, and
  prediction in ecology.
\newblock Ecology {\bf 102}:e03336.

\bibitem[{Vehtari et~al.(2017)Vehtari, Gelman, and Gabry}]{Vehtari2017}
Vehtari, A., A.~Gelman, and J.~Gabry.
\newblock 2017.
\newblock {Practical Bayesian model evaluation using leave-one-out
  cross-validation and WAIC}.
\newblock Statistics and Computing {\bf 27}:1413--1432.

\bibitem[{Yates et~al.(2021)Yates, Richards, and Brook}]{Yates2021}
Yates, L.~A., S.~A. Richards, and B.~W. Brook.
\newblock 2021.
\newblock Parsimonious model selection using information theory: a modified
  selection rule.
\newblock Ecology. 10.1002/ecy.3475 .

\bibitem[{Zhang(2008)}]{Zhang2008}
Zhang, C.
\newblock 2008.
\newblock {Prediction error estimation under Bregman divergence for
  non-parametric regression and classification}.
\newblock Scandinavian Journal of Statistics {\bf 35}:496--523.

\bibitem[{Zou and Hastie(2005)}]{Zou2005}
Zou, H., and T.~Hastie.
\newblock 2005.
\newblock Regularization and variable selection via the elastic net.
\newblock Journal of the Royal Statistical Society: Series B (Statistical
  Methodology) {\bf 67}:301--320.

\end{thebibliography}

\clearpage

\section{Supplementary Figures}
 
 \renewcommand{\thefigure}{S\arabic{figure}}

\setcounter{figure}{0}

\begin{figure}[h!]
	\centering
	\includegraphics[width=0.9\textwidth]{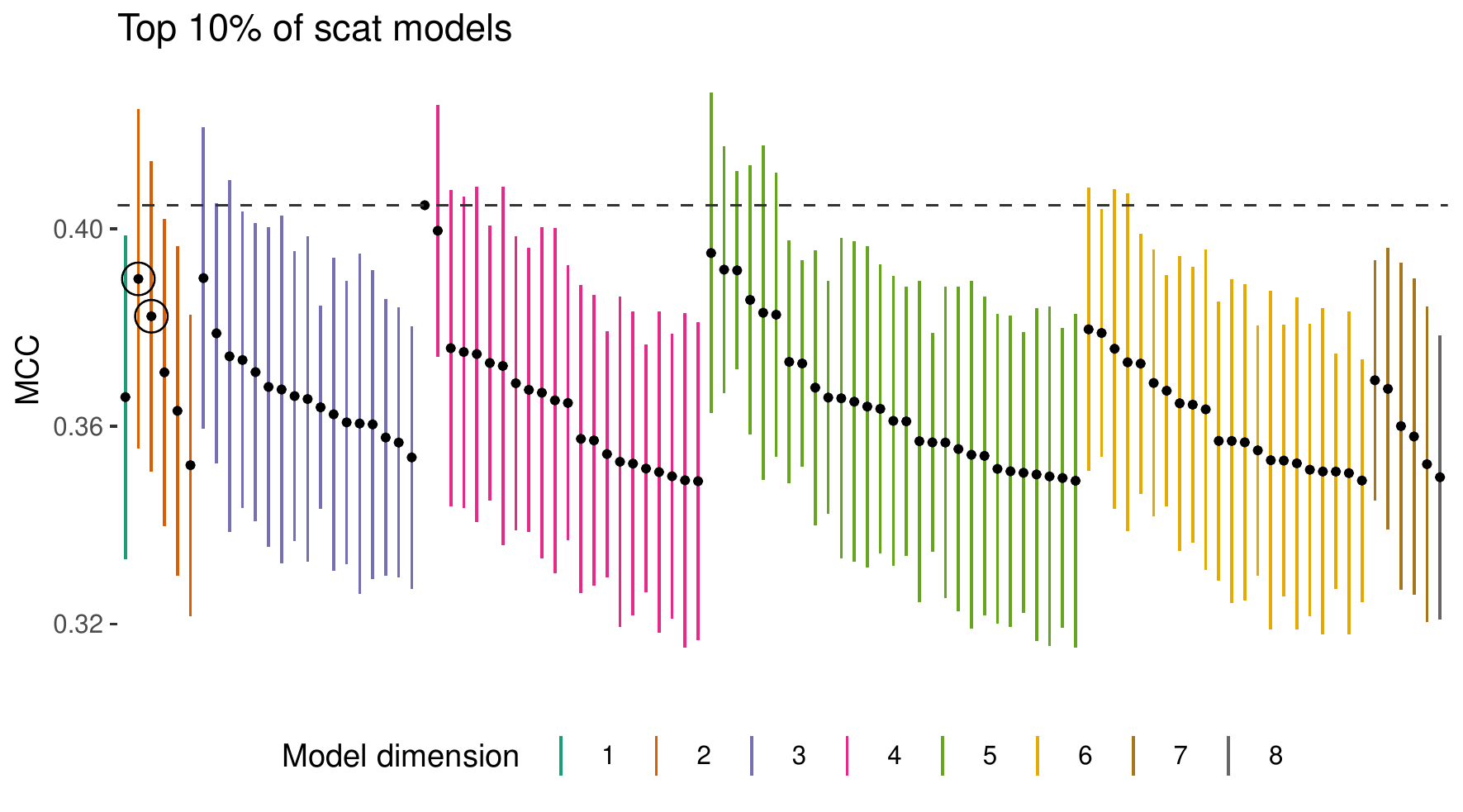}
	\caption{\label{fig:scat_modsel_top10} Model comparison using repeated 10-fold cross-validation estimates of Matthew's correlation coefficient (MCC).  The dots and bars depict the mean MCC estimate and the modified standard error \eqref{eq:se_mod}, respectively. Two models (shown circled) satisfy the selection condition for the modified OSE rule, although the best scoring of these is likely to be the preferred selection, since adding together the three distinct predictors from these models does not lead to improved performance.}
\end{figure}

\begin{figure}[!h]
	\centering
	\includegraphics[width=0.5\textwidth]{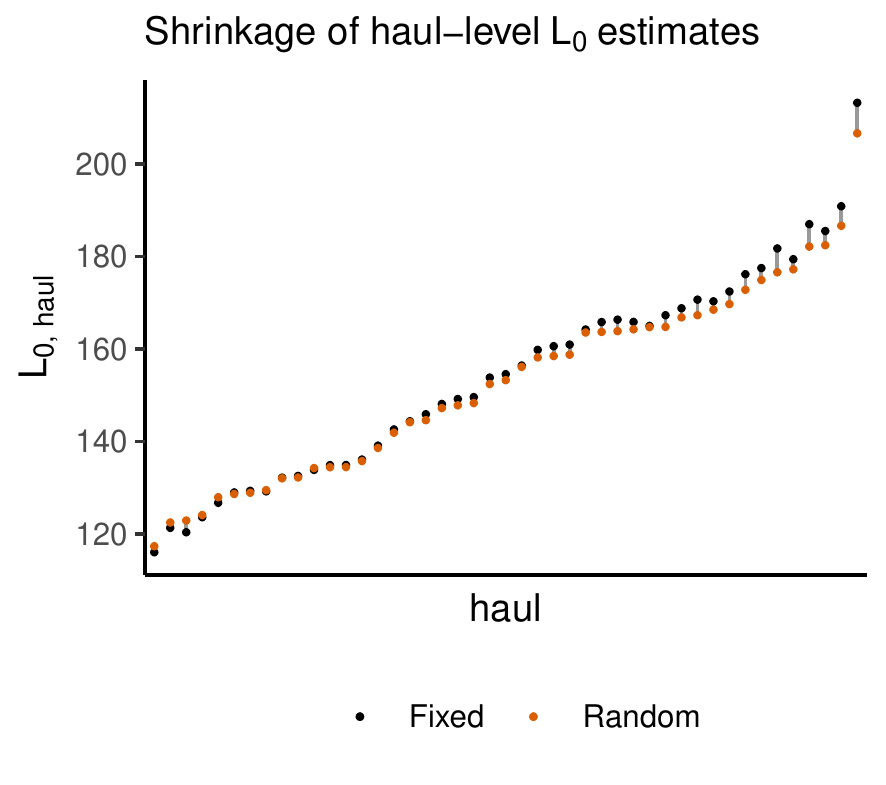}
	\caption{\label{fig:shrinkage} Haul-level length asymptote $L_0$ estimates for pinfish growth models: a comparison of fixed vs random effect estimates.}
\end{figure}

\begin{figure}[h!]
	\centering
	\includegraphics[width = 0.9\textwidth]{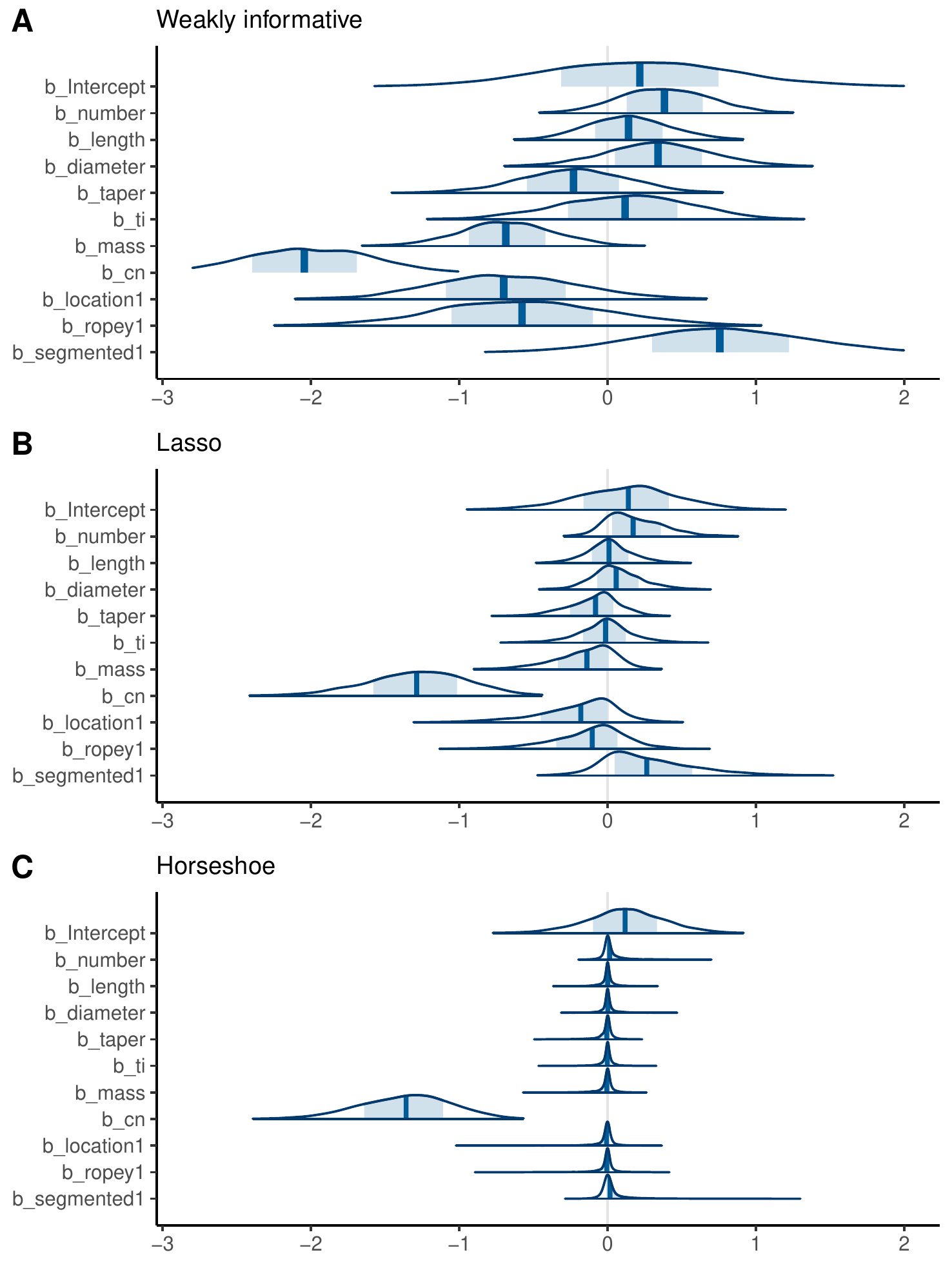}
	\caption{\label{fig:scat_reg_post} Posterior densities for the global scat model using three alternative prior specifications for model regularisation. The weakly-informative priors are Normal$(0,10)$, the LASSO priors are two-sided exponential$(1)$, and the horseshoe prior has a global scale parameter of 0.0449, following the advice of \citet{piironen2017b}.}
\end{figure}

\end{spacing}

\end{document}